\documentclass{aa}  
\usepackage{graphicx,color}
\usepackage{latexsym,amsfonts,amssymb}
\usepackage{url}
\usepackage{natbib}
\usepackage{longtable}
\usepackage{rotating}
\usepackage{xspace}
\usepackage{txfonts}
\usepackage{graphicx}
\newcommand{\CARS}{\texttt{CARS}\xspace}

\newcommand{\Elixir}{\texttt{Elixir}\xspace}

\newcommand{\SExtractor}{\texttt{SExtractor}\xspace}

\newcommand{\myarcsec}{\hbox{$.\!\!^{\prime\prime}$}}

\newcommand{\eqref}[1]{eq.~(\ref{#1})}

\begin{document}
\title{Photometric redshifts for the CFHTLS-Wide \thanks{Based on observations obtained with MegaPrime/MegaCam, a joint project of CFHT and CEA/DAPNIA, at the Canada-France-Hawaii Telescope (CFHT) which is operated by the National Research Council (NRC) of Canada, the Institut National des Sciences de l'Univers of the Centre National de la Recherche Scientifique (CNRS) of France, and the University of Hawaii. This work is based on data products produced at TERAPIX and the Canadian Astronomy Data Centre (CADC) as part of the Canada-France-Hawaii Telescope Legacy Survey, a collaborative project of NRC and CNRS.}}
\titlerunning{Photometric redshifts for the CFHTLS-Wide}
\author{F. Brimioulle\inst{1}, M. Lerchster\inst{1}\fnmsep\inst{2},   S. Seitz\inst{1}\fnmsep\inst{2},  R. Bender\inst{1}\fnmsep\inst{2}, and J. Snigula\inst{2}\fnmsep\inst{1} 
}
\authorrunning{Brimioulle et al.}        
\offprints{\\F. Brimioulle  \email{fabrice@usm.lmu.de}}
\institute{Universit\"atssternwarte M\"unchen, Ludwig-Maximillians Universit\"at, Scheinerstr. 1, 81679 M\"unchen, 		   Germany
         \and
         	   Max-Planck-Institut f\"ur extraterrestrische Physik, Giessenbachstra\ss e, 85748 Garching, Germany
             }

\date{Received; accepted}


\abstract
{}
{We want to derive bias free, accurate photometric redshifts for those
fields of the Canada-France-Hawaii Telescope Legacy Survey (CFHTLS)
Wide Data which are covered in the $u^*, g', r', i'$ and $z'$ filters
and are public on January 2008. These are 21, 5 and 11 square degrees
in the \texttt{W1}, \texttt{W3} and \texttt{W4} fields with photometric data for $1.397.545$
(\texttt{W1}), $366.190$ (\texttt{W3}) and $833.504$ (\texttt{W4}) galaxies i.e. for a total of
$2.597.239$ galaxies.}
{We use the photometric redshift code PHOTO-z of Bender et
al. (2001).}
{To study the reliability of the photometric redshifts for the CFHTLS
broad band filter set we first derive redshifts for the CFHTLS-Deep
field \texttt{D1}, and compare the results to the spectroscopic and photometric
redshifts presented in Ilbert et al. (2006).
After that we compare our redshifts for the  \texttt{W1},  \texttt{W3} and  \texttt{W4} fields
to about 7500 spectroscopic redshifts from the VVDS
therein. For
galaxies with $17.5 \leq i'_{AB} \leq 22.5$ the accuracies and outlier
rates become $\sigma_{\rm \Delta z/(1+z)} = 0.033 $, $\eta\sim$ 2 \% for
\texttt{W1}, $\sigma_{\rm \Delta z/(1+z)} = 0.037 $, $\eta \sim $ 2\% for
\texttt{W3} and $\sigma_{\rm \Delta z/(1+z)} = 0.035 $, $\eta=\sim$ 2.5 \%
outliers for \texttt{W4} fields.\\
Finally we consider the photometric redshifts of
Erben et al. (2008) which
were obtained with exactly the same photometric catalog using the
BPZ-redshift code and compare them with our computed redshifts.
For the total galaxy sample with about 9000 spectroscopic redshifts from VVDS,
DEEP2 or SDSS we obtain a $\sigma_{\rm \Delta z/(1+z)}=0.04$ and $\eta=5.7\%$
for the PHOTO-z redshifts.
We also merge the subsample with good photometric
redshifts from PHOTO-z with that one from BPZ 
to obtain a sample which then contains `secure'
redshifts according to both the PHOTO-z and the BPZ codes. This sample
contains about 6100 spectra and the photometric redshift qualities
become $\sigma_{\rm \Delta z/(1+z)}=0.037$ and $\eta=1.0\%$ for our PHOTO-z
redshifts.
}
{We conclude that this work provides a bias free, low dispersion
photometric redshift catalog (given the depth and filter set of the
data), that we have criteria at hand to select a `robust' subsample
with fewer outliers.  Such a subsample is very useful
to study the redshift dependent growth of the dark matter fluctuations
with weak lensing cosmic shear analyses or to investigate the redshift
dependent weak lensing signal behind clusters of galaxies in the
framework of dark energy equation of state constraints.\\
The PHOTO-z photometric redshift catalog is provided on request. 
Send emails to fabrice@usm.lmu.de}

\maketitle
%
\section{Introduction}
\label{sec:intro}
The CFHTLS Wide survey plans to image 170 square degrees in four
patches of 25 to 72 square degrees through the whole filter set ($u^*
g' r' i' z'$) down to $i'=24.5$.  This survey will (among other goals)
allow to study the evolution of galaxies, the large scale structures
as traced by galaxies, groups and clusters of galaxies.  Due to its
superb PSF-quality one can also directly study the line of sight
matter distribution through weak lensing analysis. Full exploitation
of the data requires the redshift of galaxies to be known in order to
obtain the 3 dimensional arrangement of galaxies and to turn the
observed galaxy colors into restframe properties. Obtaining spectra
for millions of galaxies is impossible at the moment.
The photometric redshift technique, however, can provide redshifts for
large numbers of faint galaxies with an accuracy that eg. allows
galaxy evolution studies or 3D lensing analysis.  For high quality
photometric redshifts the photometry should cover a wide wavelength
range. The necessary wavelength range has to be adapted to the depth
of the survey. For a survey as shallow as SDSS, NIR data are not
essential since basically all redshifts are low ($z<1$), and there are
hardly any SED-z degeneracies. This holds as long as U-band data
are available, which locate the Balmer or $4000$ Angstroem break;
therefore, the central wavelength of the U-band (or in general, the
bluest) filter determines the redshift above which photometric
redshifts are trustable (see Gabasch et al. 2007 and 
Niemack et al. 2008 for the impact of GALEX data on photoz-accuracies).
The CFHTLS-Wide (\texttt{W1}-\texttt{W4}) and Deep surveys are deeper than SDSS, which
implies that larger redshifts but also smaller absolute luminosities
and thus different SED-types are traced. In this situation SED-z
degeneracies can occur (eg. degeneracies between a redshift $z=0.7$
emission line galaxy and a 'normal' $z=1.2$ galaxy) which can be cured
either with bluer U-band (from space) or with NIR data. The classical
`catastrophic' failures become really relevant only in data as deep or
deeper than the CFHTLS-Deep fields.
Only in data as deep as this there is a significant number of galaxies 
with sufficiently high redshifts where 
Lyman break and $4000$ Angstrom might be misidentified. This effect
 could be supressed with absolute luminosity
priors, and (almost) avoided with NIR data.
\\
The principle disadvantage of photometric redshifts is the
relatively low redshift resolution (due to the width
of filters) compared to spectroscopic redshifts. On the other hand,
photometric redshifts have turned into a vital tool in resolving
redshift ambiguities where spectra show single (emission) line features only
(Lilly et al. 2006).\\
\\
In the Erben et al. \cite{erben08} we combined
publicly available\footnote{We are very greatful for the CFHTLS survey team to
 conduct the survey, and for the Terapix team 
(http://terapix.iap.fr/) for developing software, for preprocessing the images and carrying out the numerous data control steps. A description of the CFHTLS survey can be found at \url{http://www.cfht.hawaii.edu/Science/CFHTLS}}
$u^*g'r'i'z'$ data, remapped and coadded the
frames, derived an $i'$-band detected photometric catalog; we also made
pixel based data and the photometric catalog public. This catalog
also contains a photometric redshift estimate obtained with the
BPZ-code (Ben\'itez et al. 2000), using the original
CWW-templates and redshift priors developed from Ben\'itez for the
HDF.
The redshifts suffer a bias for $z<1.0$, where low redshift galaxies 
are at too high redshift and high redshift galaxies are at too low redshifts. 
Also, it was emphasized in Erben et al. \cite{erben08} that photometric 
redshifts in this catalog are not trustable above redshifts of $z=1.4$.
Providing redshifts with strongly reduced bias is the main goal of this paper:
In section \ref{sec:data} we briefly summarize our previous work. We then
describe the spectroscopic data that we use for calibration and for
redshift accuracy tests in section \ref{sec:spec}. We shortly describe
the photometric redshift method of Bender et al. \cite{bender01}
and our relative zeropoint
recalibration method in section \ref{sec:photozmethod}.  In section
\ref{sec:photoztest} we demonstrate that this redshift code is bias free and works 
as good as the method of Ilbert et al. \cite{ilbert06}
by comparing to spectroscopic and photometric redshift results
of Ilbert et al. for the CFHTLS \texttt{D1} field. We then
present  photometric redshifts for the \texttt{W1}, \texttt{W3} and \texttt{W4} fields, and
infer their quality (as a function of SED type and brightness) 
from spectroscopic data in these fields. 
Our redshifts are then compared to the previous ones from 
Erben et al. \cite{erben08}. We also 
empirically correct for their redshift bias. We end up with two
sets of photometric redshifts, which can also be used in combination,
to select a subsample with most reliable redshifts.
%
\section{Data Acquisition, Reduction and Photometric Catalogs}
\label{sec:data}
\begin{table}
\begin{minipage}[t]{\columnwidth}
\caption{CFHTLS Wide Fields: location}
\label{tab:fieldpos}
\centering
\renewcommand{\footnoterule}{}  
\begin{tabular}{lccccc}
\hline \hline
ID& RA     &  Dec     \\
~ &(J2000) & (J2000)  \\
\hline
\texttt{W1}  &02:18:00  &     -07:00:00  \\
\texttt{W2}\footnote{We will provide photozs for this field later}  &08:54:00  &     -04:15:00 \\
\texttt{W3}  &14:17:54  &     +54:30:31  \\
\texttt{W4}  &22:13:18  &     +01:19:00  \\
\hline
\end{tabular}
\end{minipage}
\end{table}
%
\begin{figure*}
\centering
\includegraphics[width=8.5cm]{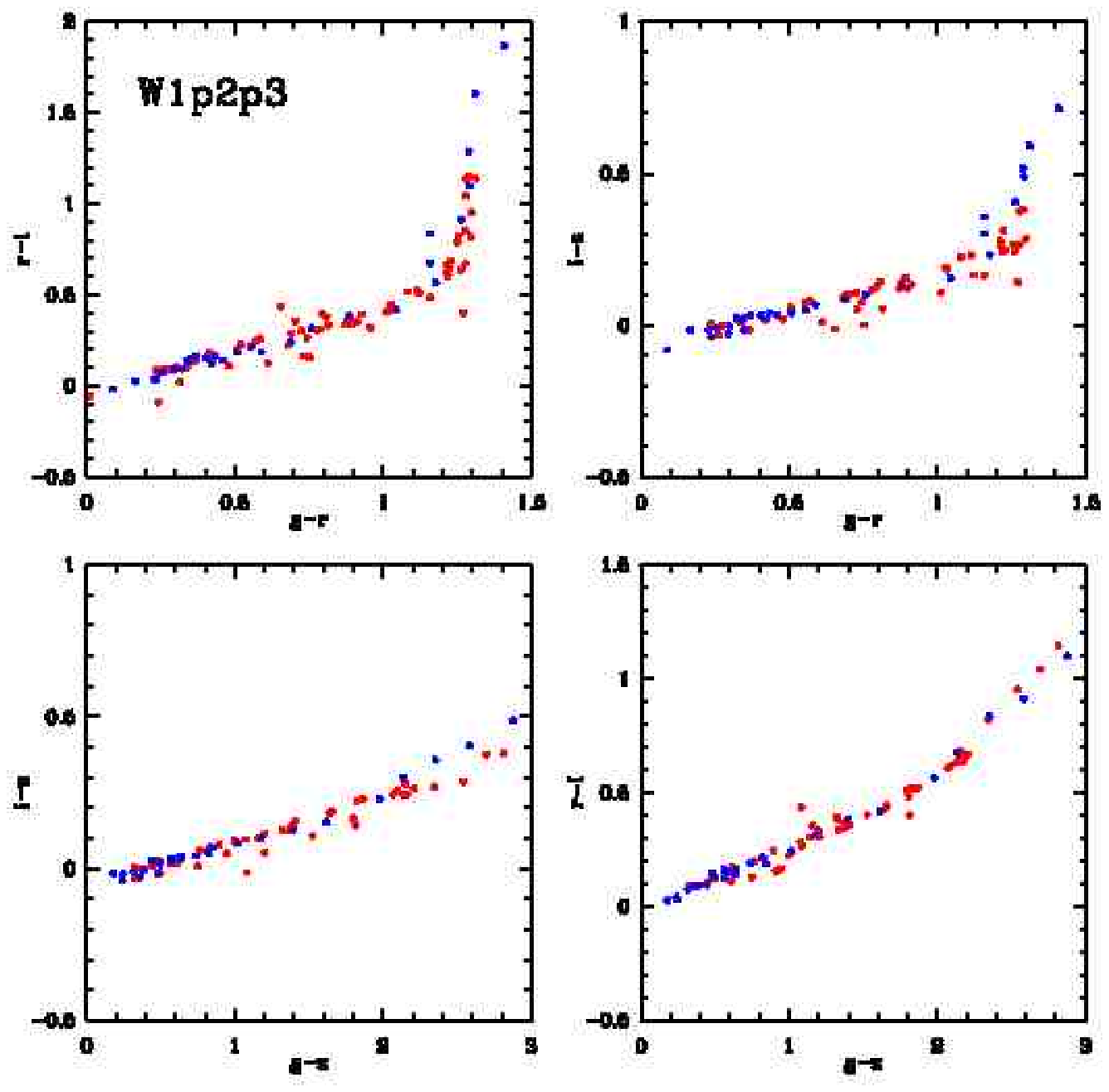}
\includegraphics[width=8.5cm]{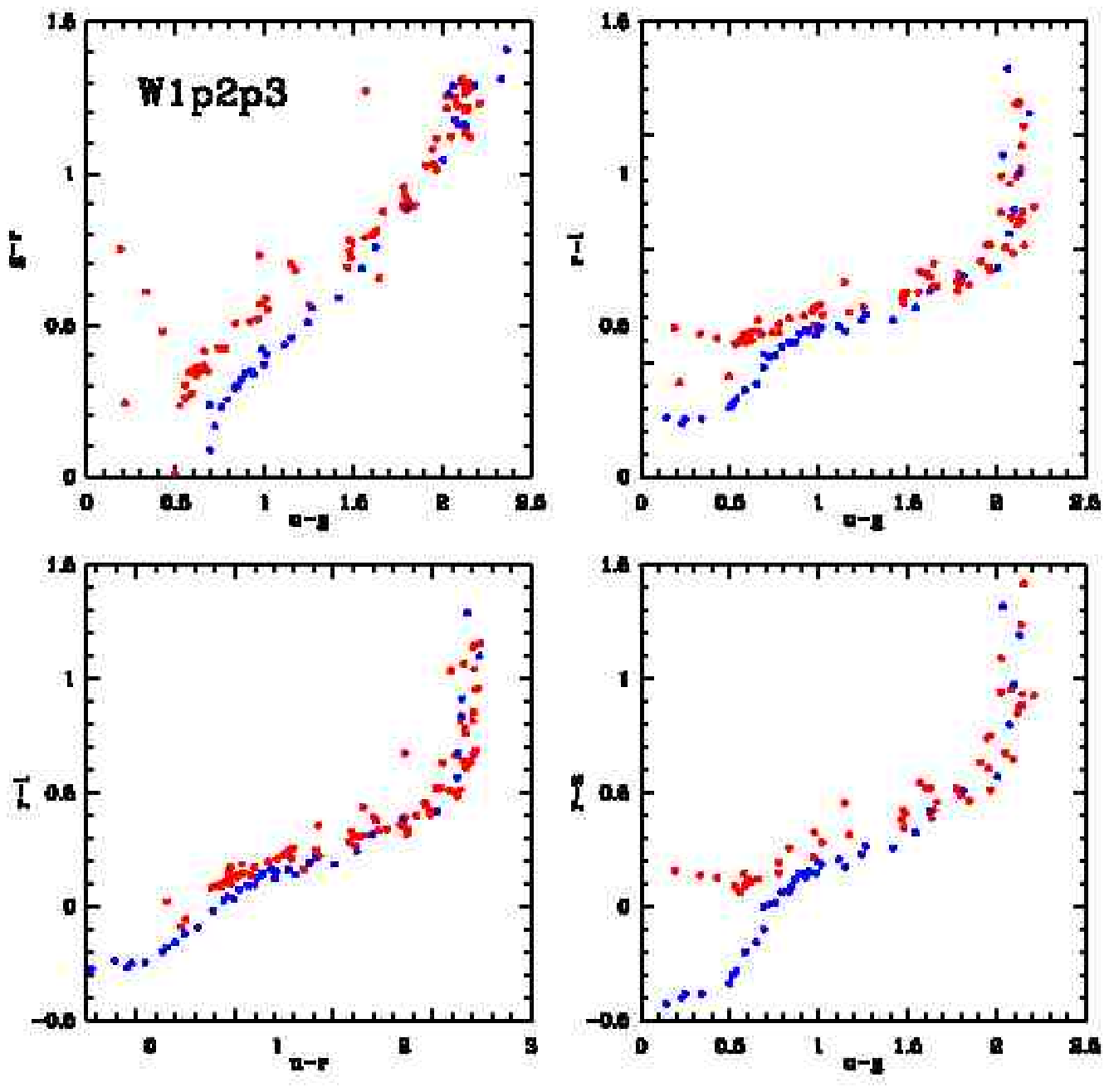} 
\caption{Color-Color diagrams of stars (red dots), plotted against the Pickles \cite{pickles98} stellar library (blue dots), on the field \texttt{W1p2p3}. Note, that all colors that do not contain a $u^*$-flux perform well. The potential reasons for the mismatch of $u^*$-band data are discussed in the text. To obtain these color-color diagrams, a significant shift ($-0.20$mags) in $u^*$-flux was applied, the shifts for the other bands ($g' r' i' z'$) were $0.00$, $0.00$, $-0.06$ and $0.13$ magnitudes.}
\label{FigW1p2p3color}
\end{figure*}
%
%
Here we briefly review the data acquisition, the data reduction steps
and the creation of the multicolor catalogs; more details can be
found in Erben et al. \cite{erben08}.\\ The data used in this
analysis are taken in the framework of the synoptic CFHTLS-Wide
observations with the MegaPrime instrument mounted at the
Canada-France-Hawaii Telescope (CFHT). See
\url{http://www.cfht.hawaii.edu/Science/CFHTLS/} and
\url{http://terapix.iap.fr/cplt/oldSite/Descart/summarycfhtlswide.html}
for further information on survey goals and survey implementation. We
consider all \Elixir processed CFHTLS-Wide fields with observations in
all five optical bands $u^*g'r'i'z'$ which are publicly
available. After downloading all data from CADC we further process
them with our GaBoDS/THELI pipeline (astrometric solution, remapping,
stacks). The stacked data have a pixel size of 0.186'', a typical PSF
of 0.8'' and limiting AB-magniudes of about 24.5 ($5\sigma$ within a
2'' aperture for a point source) in the $i'$-band. 
\\ 
For the creation of the multicolor catalogs we first cut all images in
the filter $u^*g'r'i'z'$ of a given field to the same
size. We then measure the seeing in each band and convolve all images
with a Gaussian to degrade the seeing to that of the worst band.

For object detection we use \SExtractor in dual-image mode with the
unconvolved $i'$-band image as the detection image. We measure the
fluxes in apertures in the convolved images and obtain
aperture colors. The aperture we use for photoz estimates has a
diameter of $1.86''$.  It is important to keep track of locations
which have increased photometric errors that are not accounted for in
the \SExtractor flux errors. These are: halos of very bright stars,
defraction spikes of stars, areas around large and extended galaxies
and various kinds of image reflections. Masks are automatically
generated but then finalized by human eye. These masks can also be
used as masks where shape estimates are unreliable, and they can be
obtained from Erben et al. \cite{erben08} on request.
We generate photometric redshifts for all objects. They have a
non-zero-flag (equal to the MASK value in Erben et
al. 2008) if photometry and thus redshifts (and possibly
also shape estimates) could not be trustable.  The fraction of flagged
objects/area is about 20 percent.  This is in line with conservative
flagging, e.g., in previous work (compare Ilbert et al. 2006).
%
\section{Spectroscopic Redshifts}
\label{sec:spec}
\begin{figure*} \centering
\includegraphics[width=8.5cm]{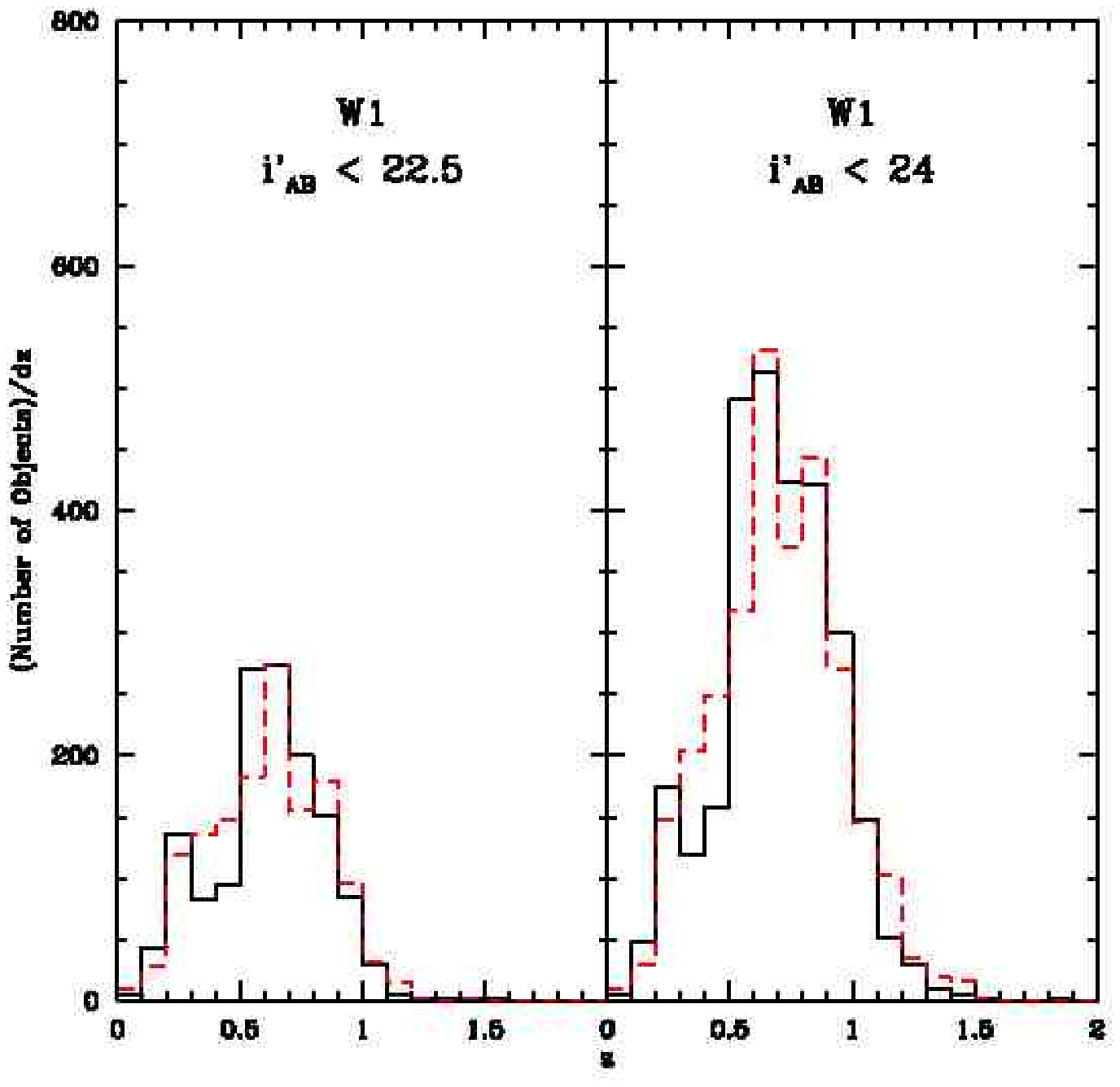}
\includegraphics[width=8.5cm]{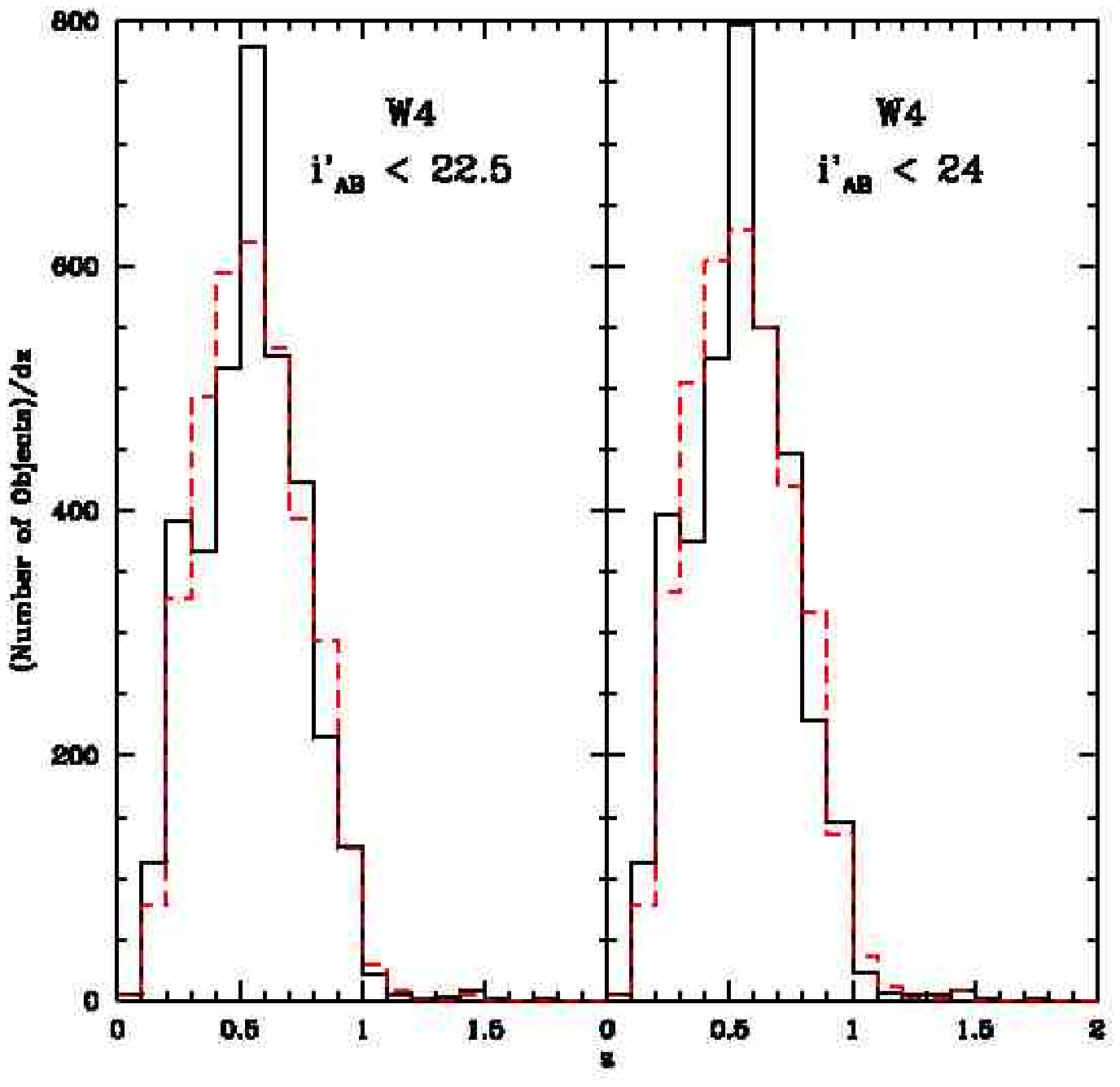}
\caption{Photometric and spectroscopic (VVDS) redshifts in the CFHTLS
\texttt{W1} (left) and the \texttt{W4} (right) fields. Left panels:
sample with $i \leq 22.5$. Right panels: sample with $i' \leq
24.5$. The black lines are for photometric redshifts and the red lines
are for spectroscopic ones, which come from the VVDS sample.  }
\label{FigHistPhotSpecComp_1_4}
\end{figure*}

The CFHTLS Wide fields  \texttt{W1} and  \texttt{W4} have a good spectroscopic coverage: Le
F\`evre et al. \cite{lefevre04} and \cite{lefevre05} have released
a catalog of 8981 spectra of galaxies, stars and QSOs with 17.5
$<$$i_{AB}$$ < $ 24.0 in the VVDS 0226-04 field (which is located
within the W1-field).  The spectroscopic redshifts are within 0 $<$ z
$<$ 5, with a median redshift of about 0.76. The sample covers 0.5
$deg^2$ of sky area.  For the CFHTLS \texttt{W4} there are 17928
spectra of galaxies, stars and QSOs (Le F\`evre et
al. 2004, 2005, Garilli et
al. 07) located in the VVDS-F22 field with a magnitude
limit of $i_{AB} < 22.5$. This sample covers 4 ${\rm deg}^2$ of sky area.\\
The online database gives access to the redshifts and quality flags,
to the multi-wavelength photometric information, as well as to the
images and VIMOS spectra. The data can be accessed via the
CENCOS\footnote{\url{http://cencosw.oamp.fr/EN/index.en.html}}web
tool.\\
The CFHTLS \texttt{W3} has public spectroscopic data from the DEEP
survey\footnote{ Data presented herein were obtained at the W. M. Keck
Observatory, which is operated as a scientific partnership among the
California Institute of Technology, the University of California and
the National Aeronautics and Space Administration. The Observatory was
made possible by the generous financial support of the W. M. Keck
Foundation.} (Davis et al. 2003, 2007; Vogt
et al. 2005; Weiner et al. 2005). The DEEP1
redshift catalog
\footnote{\url{http://mingus.as.arizona.edo/~bjw/papers/}} contains
658 objects with a median redshift of z = 0.65.  The DEEP2 DR3
redshift catalog
\footnote{\url{http://deep.berkeley.edu/DR3/zcat.dr3.v1_0.uniq.dat}}
contains 47700 unique objects with redshifts $>$ 0.7 and covers 4
regions, each $120'\times30'$ large. The targets were selected from
the CFHT12K BRI imaging, eligible DEEP2 targets have $18.5\leq
R_{AB}\leq24.1$. The region on the Groth Survey Strip, with
$120'\times15'$ has an overlap to the CFHTLS \texttt{W3} field. \\
For the comparison to photometric redshifts we consider galaxies with
trustworthy (for the \texttt{W1} and \texttt{W4} $\geq$ 95\%, for the
\texttt{W3} $\equiv 100\%$) spectroscopic redshifts only.  Due to low
S/N (at high redshift) and the limited wavelength ranges of the
spectra only 2933 objects on the CFHTLS \texttt{W1}, 410 objects on
the CFHTLS \texttt{W3} and 3688 objects on the CFHTLS \texttt{W4}
could be considered. In Fig.~\ref{FigHistPhotSpecComp_1_4} we compare
the photometric redshift distribution with the VVDS spectroscopic
redshift distribution. The agreement is very good.
\\ 
Our data from
patches \texttt{W3} and \texttt{W4} have complete SDSS coverage
\footnote{Funding for the SDSS and SDSS-II has been provided by the
Alfred P. Sloan Foundation, the Participating Institutions, the
National Science Foundation, the U.S. Department of Energy, the
National Aeronautics and Space Administration, the Japanese
Monbukagakusho, the Max Planck Society, and the Higher Education
Funding Council for England. The SDSS Web Site is
\url{http://www.sdss.org/}; },
from patch \texttt{W1} only southern pointings \texttt{W1p4m0},
\texttt{W1p3m0} and \texttt{W1p1m1} have SDSS overlap. Via the
flexible web-interface
\texttt{SkyServer\footnote{\url{http://cas.sdss.org/astro/en/tools/search/SQS.asp}}} 
of the Catalog Archive Server (CAS), we get access to the spectra catalog.
\\
 For our purpose we only consider objects clearly classified
as galaxy and a redshift trustworthy $\geq$ 95\%. By matching the SDSS
catalog with our photometric catalog we end up with 39 objects on the
CFHTLS \texttt{W1}, 180 objects on the CFHTLS \texttt{W3} and 309
objects on the CFHTLS \texttt{W4}. In total there are 528 objects with the
spectroscopic redshift from the SDSS (Adelman-McCarthy et al. 2007).
%
%
%
\begin{figure*}
\centering
\includegraphics[width=8.5cm]{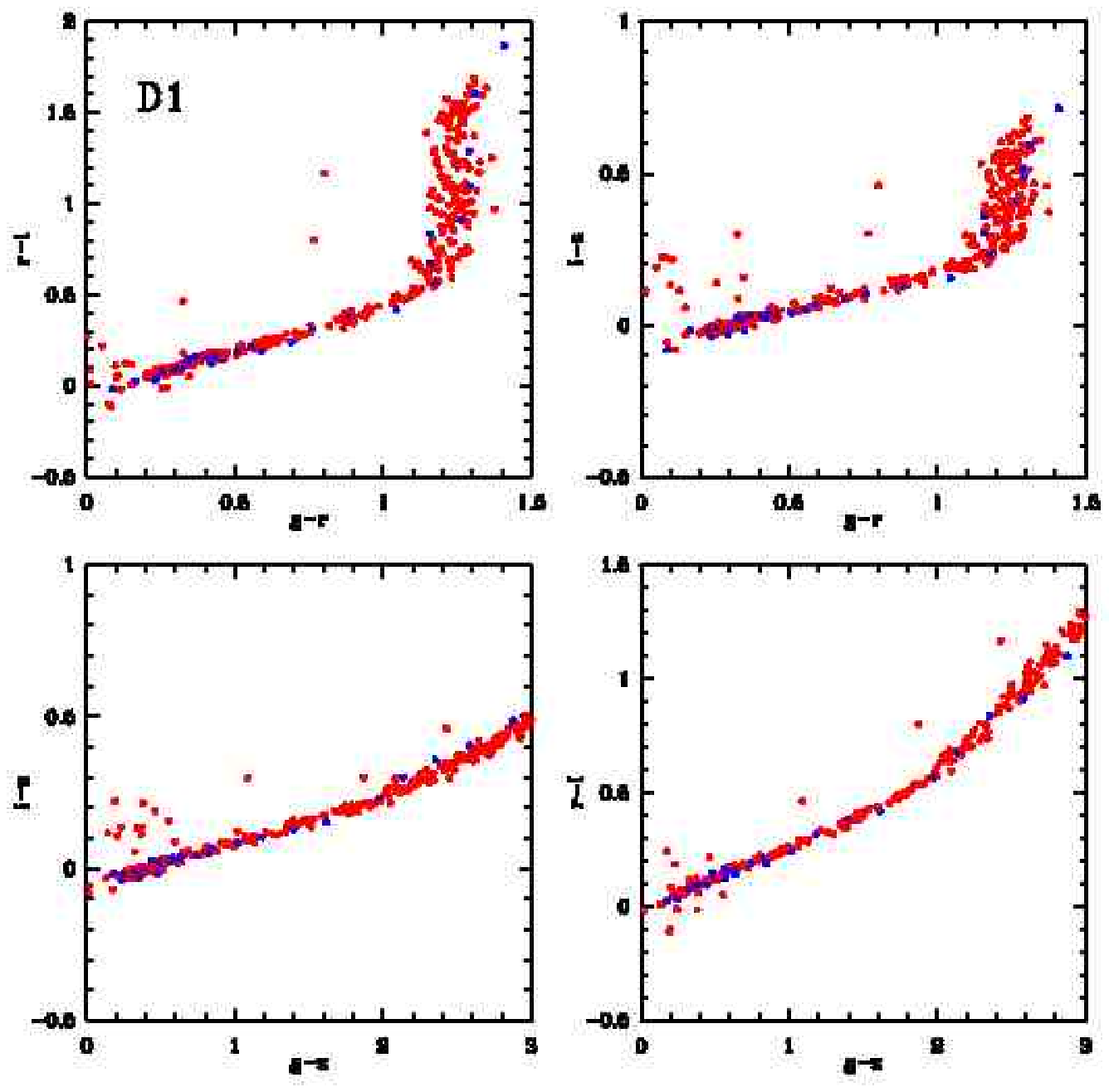}
\includegraphics[width=8.5cm]{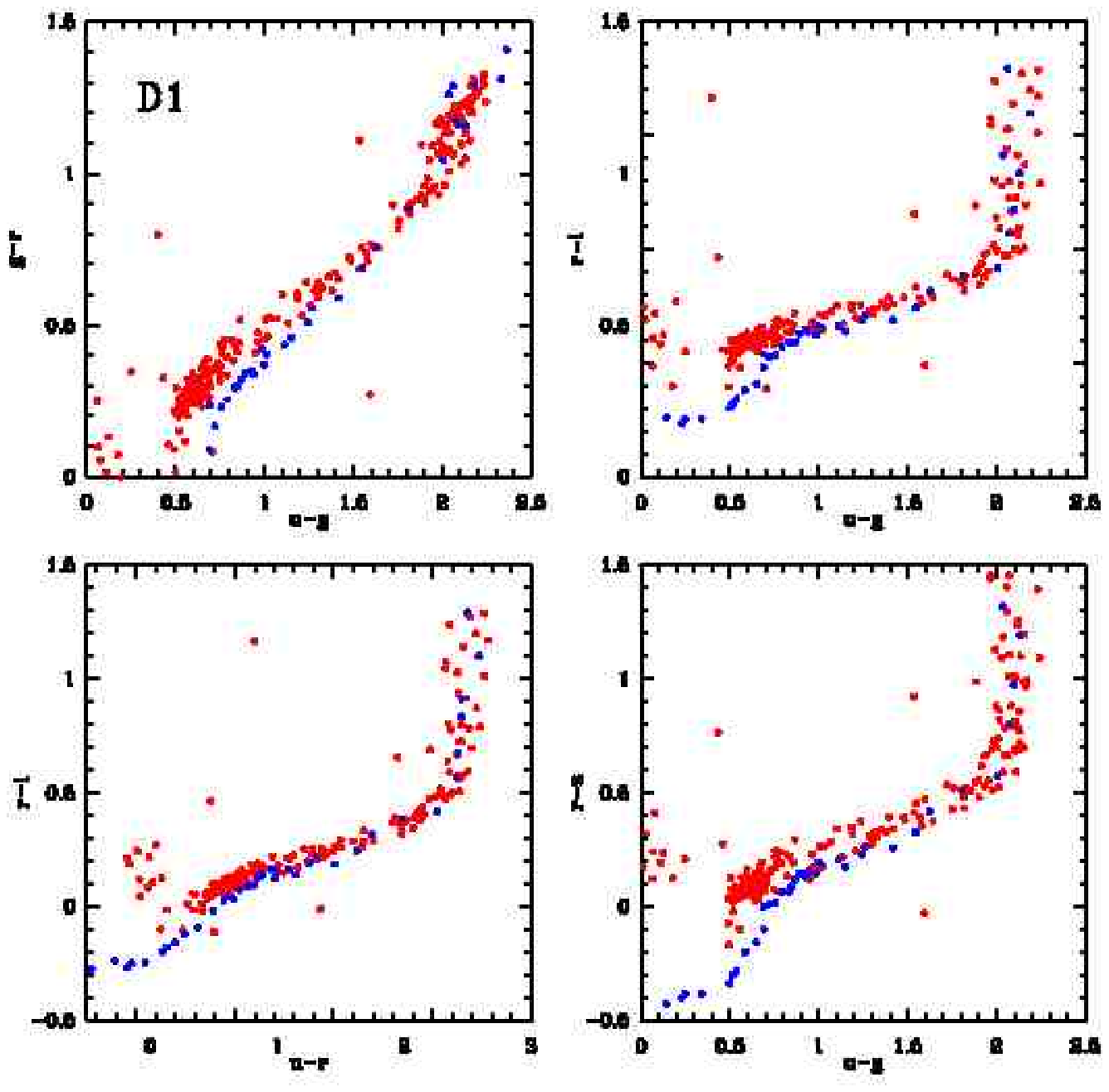} 
\caption{CFHTLS \texttt{D1} field: color-color diagrams of stars (red dots), plotted against the Pickles \cite{pickles98} 
stellar library (blue dots), on the field \texttt{D1}. Note, that all colors that do not contain a $u^*$-flux perform well. 
To obtain this color-color diagram, a significant shift ($-0.25$mags) in $u^*$-flux was applied, the shifts for the other bands ($g' r' i' z'$) were $-0.06$, $0.00$, $-0.10$ and $0.04$ magnitudes.}
\label{FigD1color}
\end{figure*}
%
%
\section{Photometric redshift method and photometric calibration of 
CFHTLS Wide and Deep data}
\label{sec:photozmethod}
We use the PHOTO-z code of Bender et al. \cite{bender01} (see also
Gabasch et al. 2004 about the construction of SED templates). The code
calculates for each SED the full redshift likelihood function
including priors for redshift and absolute luminosity. The stepsize
for the redshift grid is equal to $0.01$. 
Our aim in this paper is to estimate photometric redshifts from the 
optical bands of the CFHTLS-Wide data
based on approved SED-templates and prior settings from earlier publications.
The SEDs and priors used for this work are the same as in Bender  
et al. \cite{bender01} and Gabasch et al. 2004.
Development of new SED-templates adapted to the CFHTLS data, application of 
better adjusted priors and further investigations, data from near 
infrared bands will be part of an upcoming paper (Lerchster et al. in prep.).
Generally, our priors are
weak and do hardly influence the result. 
The luminosity prior deals the fact that the absolute luminosity 
of galaxies are limited and do not exceed a certain value. It is flat over a broad
range of restframe luminosities with a supression of absolute Magnitudes brighter 
than -25 and fainter than -13 by a factor of 2 in probablilty at these luminosity values.
The relatively strongest prior is the redshift prior for old stellar populations. 
The redshift prior considers that certain SED-Types do
not exist at higher redshifts, it supresses e.g. the probability for elliptical 
galaxies at a redshift of 0.8 by a factor of 2.
We choose a
prior which makes red SED types at $z=0.6$ and S0-like galaxies at
$z=1$ one fifth as likely as at $ z=0.1$.  The redshift priors for
other SED types are almost flat.  The redshift of the highest
probability among all SEDs becomes the `photometric redshift' of the
object. The redshift `error' of the object is obtained as
\begin{equation} \triangle z_{\rm phot}=\sqrt {\sum_{i,j} (z_i -
z_{max})^2 \cdot P_{ij}},
\end{equation}
where the sum runs over all (discrete) redshifts $z_i$, and all SEDs
and $P_{ij}$ is the contribution of the j-th SED to the total
(normalized) probability function at redshift $z_i$.  Hence, the
meaning of the `error' is how well the galaxy is locatable in redshift
space around the `best' redshift. Sometimes, the redshift
probabilities have a maximum at another distinct redshift (where so
called `catastrophic outliers' could arise). The potential
redshift-SED-degeneracy can be read off in the chi-squares of the most
likely and the second most likely SED (which in general is different from
the redshift likelihood ratios).
 
They could be improved with the now available,
much larger spectroscopic and photometric datasets. This is subject
of future work and will allow to further improve the photometric redshift
accuracy.
\\
The photometric calibration of the data is very important.  Small
errors in the photometric zero-points or false assumptions on the
wavelength dependent transmission of the system (sky, telescope
optics, filters, CCDs) have to be avoided or corrected. As throughput
of the system we use the filter curves in
\url{http://www1.cadc-ccda.hia-iha.nrc-cnrc.gc.ca/community/CFHTLS-SG/docs/extra/filters.html}.
These include optics (wide field corrector, image stabilizing plate,
camera window), the mirror (approximated with the reflectivity of
freshly alumium coated glass) and the CCDs (QE is given only between
350 and 1000 nm).  For the atmospheric extinction we used
\url{http://www.cfht.hawaii.edu/Instruments/ObservatoryManual/
om-extinction.gif} (for the blue optical part) where extinction is
featureless. This wavelength dependent extinction shifts the effective
wavelength of the filters to the red (relevant for the $u^*$-band,
where it implies a shift of about 20 Angstroems.
\\
With these ingredients we calculate the location of the Pickles
stellar libary stars (Pickles 1998) in color-color
diagrams and compare them with colors of observed stars.  The observed
comparison stars are selected in the central region of the frames
using their \SExtractor CLASS\_STAR and SExtractor flag parameters. We derive
their aperture colors (after seeing matched convolution).  An example
is shown in Fig.\ref{FigW1p2p3color}: red dots are used for measured
stellar colors, and blue dots are used for the Pickles libary stars.
Stellar sequences are well located and can evidently be used to 
measure relative zeropoint offsets. 
\\
Our  $u^*-r'$ vs $r'-i'$ 
diagram looks very different from that of Erben et al. \cite{erben08},
where the stars show a huge scatter in $u^*-r'$ colors for large values of 
$r'-i'$. We show only the brightest (unsaturated) 
stars with highest \SExtractor CLASS\_STAR 
values, which implies that photometric errors are very small
for stars in our diagrams. This then allows, to measure relative zeropoint 
offsets and to see whether the spread of stellar colors and the 
shape of the color-colors diagrams also agrees with expectations from stellar 
libaries for the assumed system throughput.
\\
The $r'-i'$ vs $g'-r'$ diagram of observed and libary stars has a very
strong curvature at $g'-r'=1.2$. This implies, that one can adjust the
zeropoint offsets of the $g'$ {\it and} $i'$-bands relative to the
$r'$-band very well. If this is done one can proceed with the
$z'$-band, using, e.g. the $g'-z'$ colors.  In this way one gets
zeropoints, for which stars have colors consistent to the Pickles
libary. After these adjustments g', r' and i' band data usually match
the expected curve very well. For $z'$-band data one might expect
larger `scatter' around the Pickles points, because some fields do
show considerable amount of fringes, which can lead to systematic
(relative to the fringe pattern) magnitude offsets for some of the
stars.
%
\begin{figure*} \centering
\includegraphics[width=8.5cm]{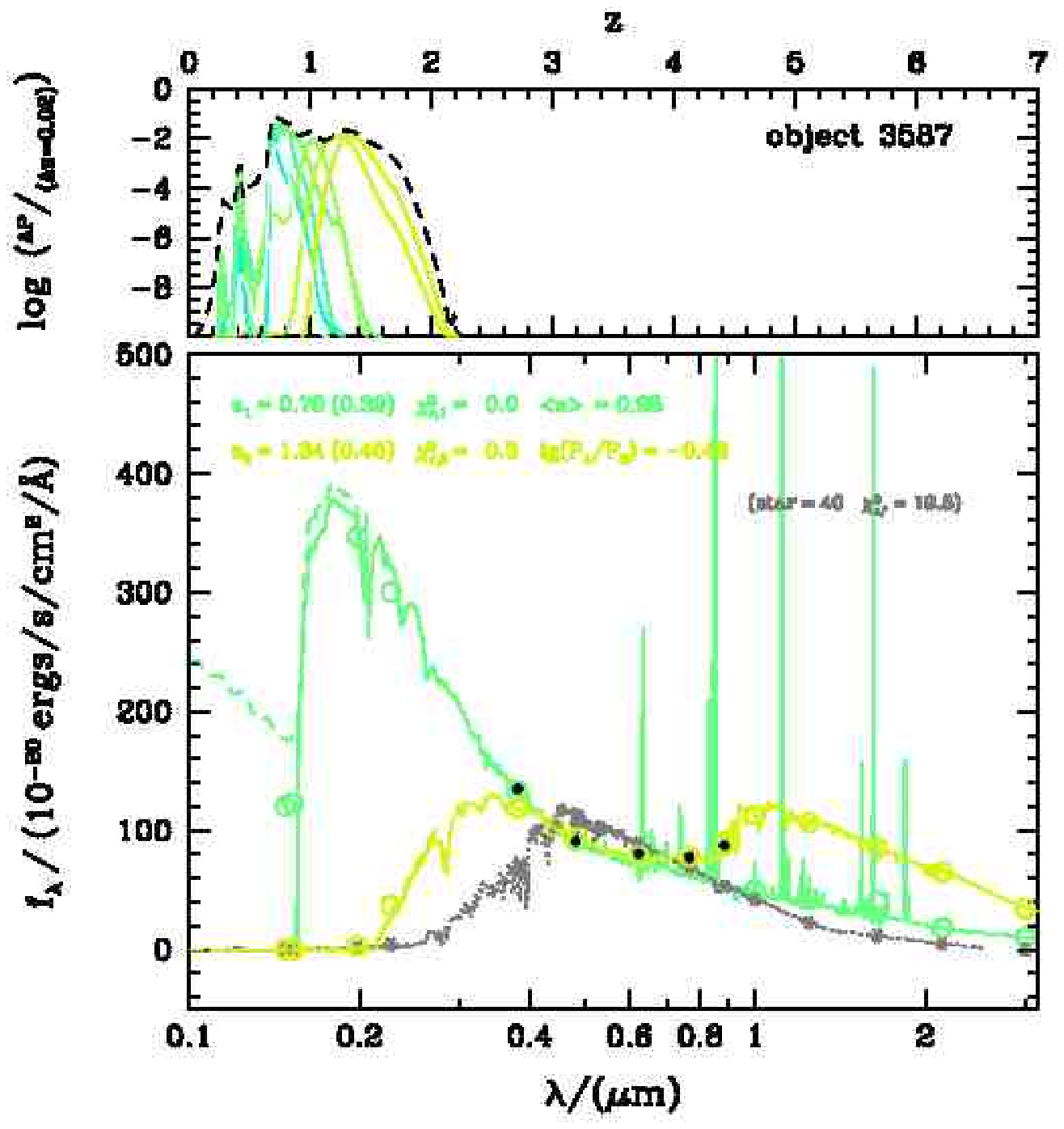}
\includegraphics[width=8.5cm]{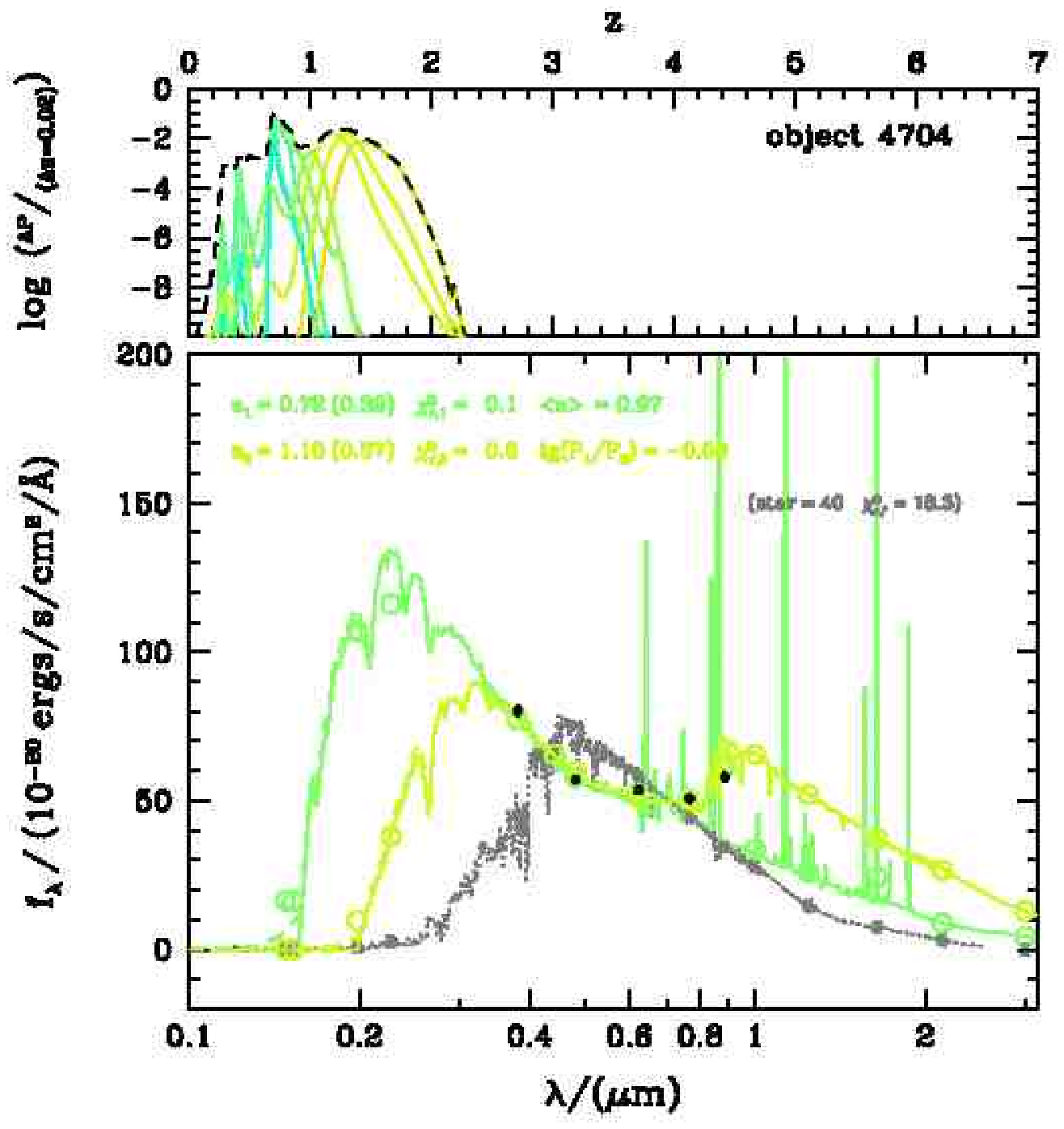}
\caption{Two examples for the SED-redshift degeneracy if optical bands
only are used: One cannot dissentangle a `normal' galaxy at $z\approx
1.2$ from a heavily starforming blue galaxy at $z\approx 0.7$ which
contributes significantly with its emission lines in the z-band. These
`blue' galaxies are faint and occur only in relatively deep data. 
The spectroscopic redshifts of the galaxies
in the left and right panels are $z=1.363$ (obj3587)
and $z=1.175 $ (obj4704). }
\label{FigSEDzdeg}
\end{figure*}
%
We finally also show all examples which involve $u^*$-band data.  They
don't match the Pickles stars, since they occupy a larger interval in
the $u^*-g'$ color (which can't be fixed by a zeropoint offset,
obviously). Since the $g'-r'$ color is corrected for zeropoint offsets
already, it means, that stars, which are blue in $g'-r'$ are bluer
than predicted in $u^*-g'$ for $u^*-g'<1.7$ than expected.  This could
be explained by the fact that stars targetted by CFHTLS are more metal
poor than those in the solor neighbourhood
and therefore bluer in the UV
. Alternatively, if
throughput is the explanation, it would surprisingly imply that the
throughput in the blue parts of the $u^*$-band has been
underestimated; this does not appear very likely.  On the other hand,
if the mismatch of colors is due to metal poor stars (UV-excess), then
the effect might have shown up more strongly in the deeper FORS Deep
Field (Gabasch et al. 2004), where more of the halo is traced, and
where the U-band filter curve is considerably bluer. As long as one
cannot firmly identify the reason for this strech in the $u^*-g'$
color, one could claim that it is not obvious, whether a `good'
stellar color-color diagram involving $u^*$-band data should match the
Pickles colors at the blue or red end (or somewhere else).
Furtheron difficulties in measuring the transmission function of 
the CFHT-$u^*$ are well known, as several versions of the transmission curve 
can be found in the web which additionally complicates the analysis. 
The effect described above shows up in any CFHT
$u^*$-band data we investigated (DEEP and WIDE fields), independent of the 
applied reduction pipeline but not in other fields we previously investigated 
(e.g. ESO DPS, in particular GOODS-S, compare Gabasch et al. 2004). 
Nevertheless, any remaining 
systematic effects leading to systematic errors in the zeropoint offset 
determination can be detected and corrected by calibrating the zeropoints 
using spectroscopic redshifts.
We therefore looked into subfields with spectroscopic data (we took
\texttt{W1p2p2, W1p2p3, W4m0m0, W4m0m1, W4m0m2, W4p1m0, W4p1m1, W4p1m2,
W4p2m0, W4p2m1, W4p2m2}) and investigated how the color-color diagrams
(involving $u^*$-band) looked like when photometric and spectroscopic
redshifts matched well.  It turned out, that a slight shift to the
color-color diagrams of stars relative to that which matches the stars
at the red end and which is shown in Fig.~\ref{FigW1p2p3color} was
necessary, for all the fields, and that this shift was consistent in
size from field to field.  We took that as a description how observed
stars have to look relative to the Pickles libary, and calibrated
other fields without spectroscopic data like that.  We have tested how
good this empiric photometric ZP-calibration works in the fields where
we predicted the photometric redshifts based on the `ideal empirical
color-color diagram' and compared to available spectroscopic data
(this was done for the fields \texttt{W3m1m2, W3m1m3} using DEEP2 spectra and
for many subfields of \texttt{W1}, \texttt{W3} and \texttt{W4} using SDSS redshifts).
%
\begin{figure*} \centering
\includegraphics[width=14cm]{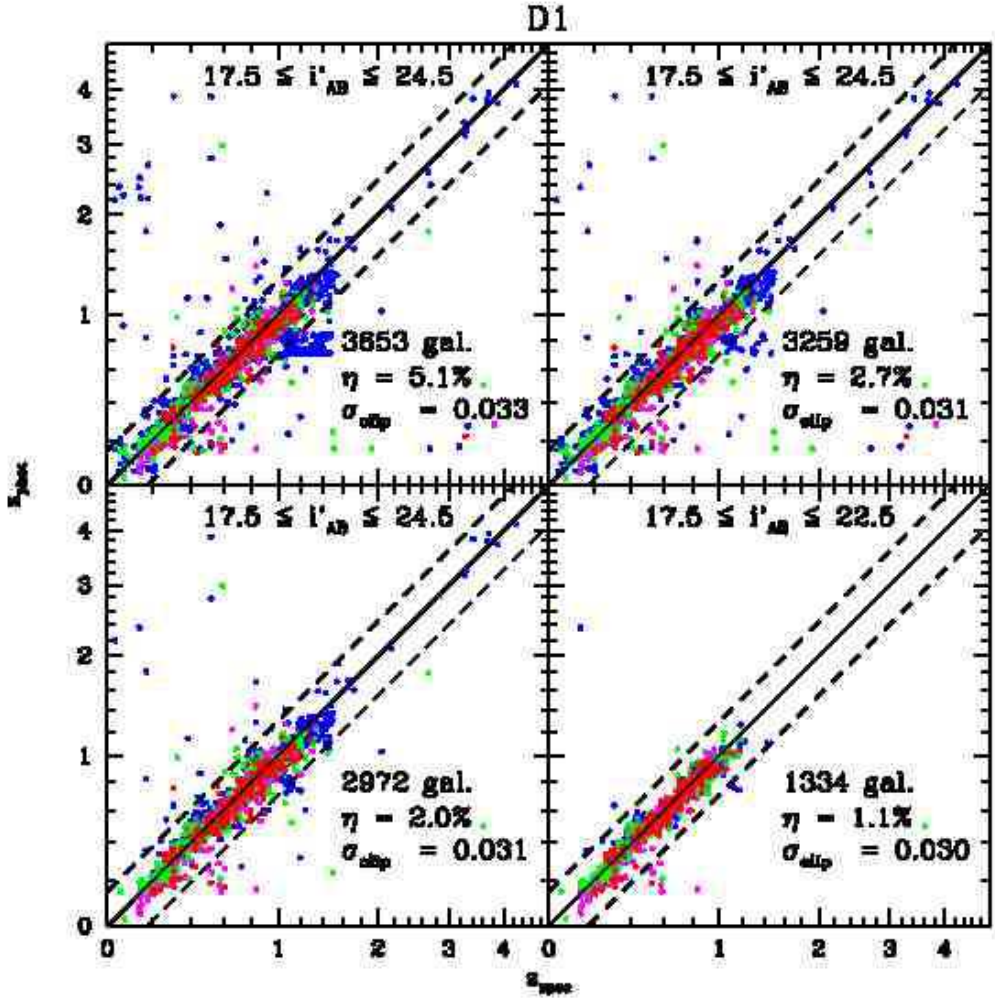}
\caption{PHOTO-z photometric redshifts against spectroscopic redshifts
for the CFHTLS \texttt{D1} field. Left upper panel: all non-stellar
objects (stars are excluded by morphology and SED-fit).  Upper right
panel: All objects from the left, but without most likely 
SEDs corresponding to starforming galaxy with strong emission lines, which
this excludes objects where a SED-photoz degeneracy is known 
(see text for more details).
Left lower panel: 
as above, but now also excluding also those objects where the width of the
redshift probability distribution at the most likely redshift is
large, $\triangle z_{\rm photoz}>0.25 * (1+z_{\rm photoz})$. 
Right lower panel: as before, but restricting the object sample to
$17.5\leq i'_{AB}\leq 22.5~mag$.
The color coding is: red symbols for SEDs that describe elliptical galaxies and S0s, blue symbols for blue, strongly star forming galaxies, the magenta and 
green coded objects are for SEDs that form a continuous sequence between the
red and blue galaxies in terms of color and star forming activity.
}
\label{FigPhotSpecCompD1}
\end{figure*}
%
%
%
\section{Photometric Redshifts: Accuracy tests with the CFHTLS \texttt{D1} data}
\label{sec:photoztest}
\subsection{Results for the complete sample}
We had been testing our photoz-method previously on the FDF, GOODS,
MUNICS and COSMOS fields 
(Gabasch et al. 2007, 2006, 
2004, Feulner et al. 2006, 
2005; Gabasch et al. 2004 and 
Drory et al. 2001).
Since any photometric redshift method is likely to fail if the system
throughput is misunderstood, we first test our method on the
CFHTLS \texttt{D1}  subfield, which has many spectra from the VVDS 0226-04
field, and where photometric redshifts have been derived by Ilbert et
al. \cite{ilbert06} before.  For the photometry of this deep field
we use the $u^*g'r'i'z'$ science data from the T0003-release (including
their weight and mask frames)\footnote{We would like to thank the
terapix team for reducing and releasing these data.}.  We then obtain
photometric catalogs in the same way as described before.  For the
spectroscopic sample we use all matchable spectra which fulfill the
redshift quality criteria defined in Section 3.  The match of
spectroscopic and photometric redshifts is quantified the same way as
in Ilbert et al. \cite{ilbert06}, i.e., we define the outlier
fraction and redshift accuracy of non-outliers as:
\begin{equation} \eta= {\rm fraction \; of \; outliers \; with } \;
|z_{spec}-z_{phot}|/(1+z_{spec})> 0.15
\end{equation}
\begin{equation} \sigma_{\rm \Delta z/(1+z)} = 1.48 \times median (|z_{\rm spec}-z_{\rm phot}
|/(1+z_{\rm spec}))_{\rm non-outliers}.
\label{sigclip}
\end{equation}
This definition of $\sigma_{\rm \Delta z/(1+z)} $ is quite different from the 'true'
dispersion
\begin{equation}
\sigma = 
\sqrt {\sum_{i}^{N_{\rm spec}}
(z_{i, {\rm photz}} - z_{i, {\rm spec}})^2 / (N_{\rm spec}-1) }
\quad,
\label{sigtrue}
\end{equation}
since it describes only the typical redshift deviation within a narrow
range around the true value. We recommend to also consider $\sigma$ in
parallel to $\sigma_{\rm \Delta z/(1+z)}$ and $\eta$, since this tells how `off'
outliers typical are. Also, if one compares the performance of
photometric redshifts with optical data alone to the case where
eg. NIR data are added, it usually happens that $\eta$ and $\sigma$
decrease, whereas $\sigma_{\rm \Delta z/(1+z)}$ can even increase, since more
data points end in the `almost true' section, but the median deviation
within this $|z_{spec}-z_{phot}|/(1+z_{spec})< 0.15$ interval can
increase. Nevertheless, one would, for many application prefer the
situation with reduced $\eta$ and $\sigma$, even if $\sigma_{\rm
clip}$ is slightly increased.
Using these definitions from above,
our fraction of outliers for the \texttt{D1} field is 
$\eta \sim 5\%$ and the accuracy
becomes $\sigma_{\rm \Delta z/(1+z)}=0.033$. This accuracy has to be
compared to that of Ilbert et al. \cite{ilbert06}. They have
obtained redshifts with the LePHARE code using about 70 template SEDs,
which have been optimized with spectroscopic-photometric data in the \texttt{D1}  field.
The photometry comes from the CFHTLS Deep1
with integration times of about 11h, 7h, 17h, 37h and 17hours in
the $u^*g'r' i'z'$-filters and PSFs between 1.1 and 0.9
arcseconds (as released in the Terapix T0003 release) and the BVRI-VVDS
with integration times of 3 to 7 hours in the CFH12K BVRI filters
and median PSF of 0.8'' -0.9'' (as described in McCracken et al. 2003).
The 50 percent point source completeness is at AB magnitudes of 26.5,
26.4, 26.1, 25.9 and 25.0 for the $u^*g'r'i'z'$-filters and
26.5, 26.2, 25.9 and 25.0 for the $BVRI$-filters according to Ilbert
et al. \cite{ilbert06}.  So, their VVDS data are fairly deep in the B and V
filters, and we therefore expect a gain in the photometric redshift
accuracies, when these data are used: The VVDS B- and V-filters can
sample breaks that are within the very broad CFHTLS-g' filter, and
the R-band helps to locate breaks that are within either the CFHTLS-r'
and i'-filters.  
In addition to the optical data Ilbert et al. \cite{ilbert06} could use deep J and K
band data for 13$ \% $ of their objects. Their results for \texttt{D1} 
are released at at \url{http://terapix.iap.fr/rubrique.php?id_rubrique=227}.
They reached values of $\sigma_{\rm \Delta z/(1+z)}=0.029$ and $\eta= 3.8\% $ 
 using all their photometric data and galaxies with 
$i<24$ according to their paper. This result is 
surprisingly close to our result if one accounts for the denser
wavelength coverage and the fractional coverage with NIR data.
\subsection{Creating subsamples with higher photo-z precision}
%
\begin{figure*} \centering
\includegraphics[width=16cm]{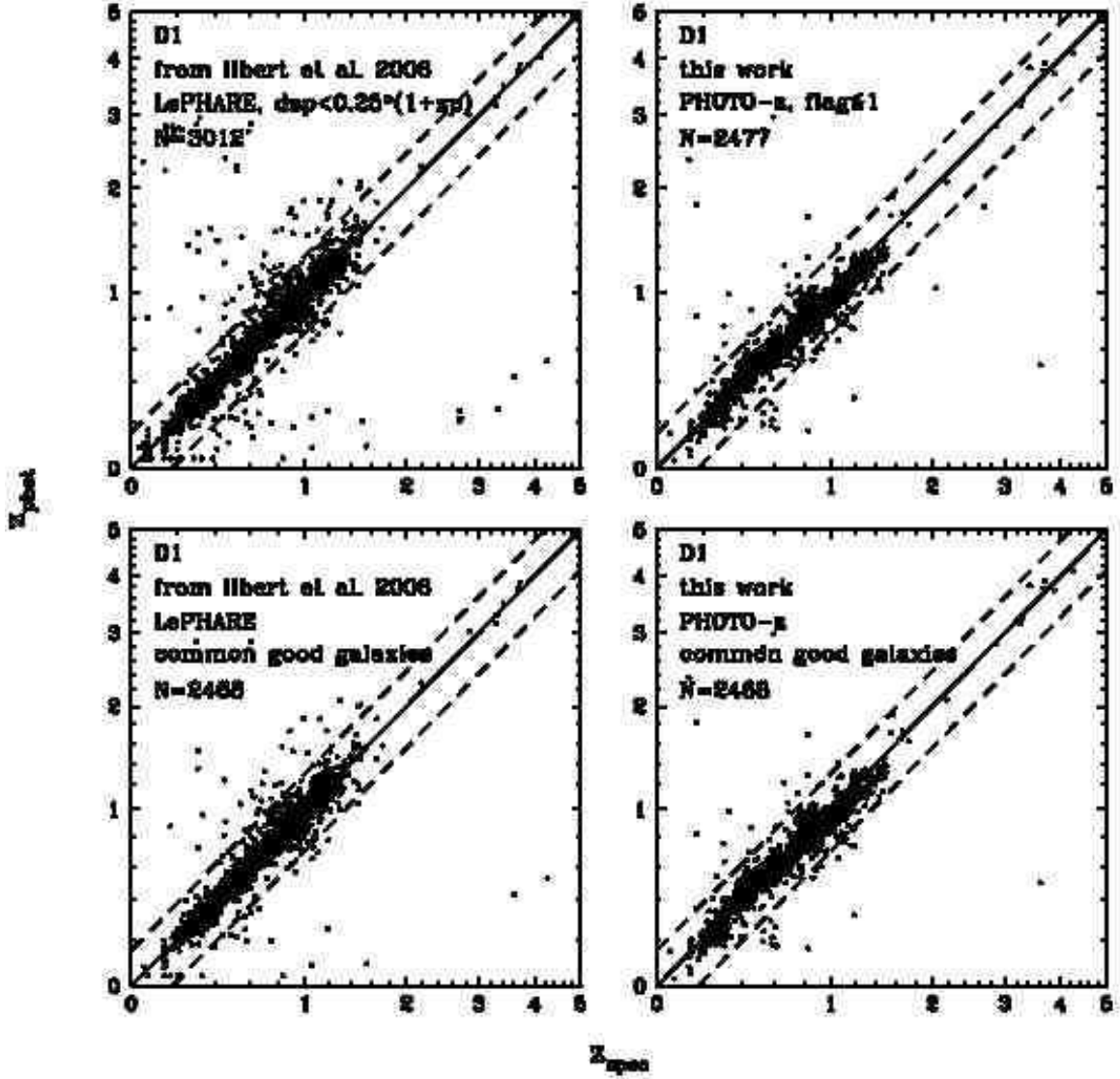}
\caption{This Figure compares the photometric redshifts derived by
Ilbert et al. (2006) with LePHARE (two left panels) with those derived
with PHOTO-z in this work (two right panels). The upper plot shows the
photometric redshifts for objects which are characterized as `good'
($
dzp:=\rm {z_{\rm sup-{68}}- z_{\rm inf_{68} } } \le 0.25(1+z_{phot}) $) 
in the LePHARE catalog and 
($\triangle_{z{\rm photoz}} <0.25 *(1+z_{\rm photoz})$ $\&$ excluding strongly starforming SED-types)  
in the PHOTO-z catalog. The lower two panels show photometric
 redshifts for those objects which are both in the `good' LePHARE 
and `good' PHOTO-z sample. }
\label{FigLePHARE_PHOTO-z}
\end{figure*}
%
%
We now explore whether we can identify a subsample of galaxies with
more precise photometric redshifts.  One expects that the quality of
photometric redshifts
 varies with the SED properties of the galaxies.  Using
optical bands ($u^*g'r'i'z'$) only, there is a redshift range where
strongly starforming galaxies cannot be discriminated from `more
normal' galaxies at other redshifts.  In Fig.~\ref{FigSEDzdeg} we show
the photometric redshift probability distribution for two
spectroscopic objects, and the SED-fit for the most likely and second
most likely SED. One can see that without near infrared data (e.g., J
or H-band) the SED-photoz degeneracy cannot be broken. In both cases
the true redshift is that of a `normal' galaxies, without strong
emsission lines. Since at the depth of the survey we do not expect
very many strongly starforming galaxies anyhow, we decided to reject
galaxies for which the strongly starforming SEDs is formally most
likely. In this way we get rid of the `systematic arms' in the spec-z
photoz plots; the accuracy becomes $\sigma_{\rm \Delta z/(1+z)} \approx 0.031$
and the outlier rate is $\eta =2.7\%$. If we further exclude objects
with photometric redshift errrors $\triangle_z{\rm photoz} >0.25 *
(1+z_{\rm photoz})$ the remaining accuracy and fraction of
outliers become $\sigma_{\rm \Delta z/(1+z)}=0.031$ and $\eta \sim 2\%$. The
fraction of galaxies that we loose with these selection criteria (small
width of the most likely redshift, and excluding galaxies with SEDs
degenerate to strongly starforming ones) is about 20 percent. We
nevertheless provide photozs for all redshifts and flag galaxies with
likely imprecise redshifts. The meaning of the photometric redshift flag
values can be looked up in the Table \ref{tab:keys} in the Appendix.
\\
Finally, if we limit
the magnitudes of catalog to $i'_{AB}\leq 22.5$~mag, which
corresponds to the limiting magnitude of the primary targets of the 
spectroscopic survey on the VVDS-F22 field, 
this leads to an accuracy of $\sigma_{\rm \Delta z/(1+z)}
=0.030$ and the fraction of catastrophic outliers is $\eta \sim1\%$.
\\
We now more directly compare the performence of our photometric
redshifts with that of Ilbert et al. \cite{ilbert06}.  We retrieve their photometric
redshift catalog \footnote{We would like to thank the authors for
providing their photometric redshift catalog to the public.}, and
merge our spectro-photometric-redshift sample with their photometric
redshift.  This `merged' sample is now smaller than our
spectro-photometric-redshift sample alone, since we skip sources that
have more than one photometric-redshift counterpart within the search
radius in the Ilbert et al. \cite{ilbert06} catalog.
\begin{table*}
\centering
\renewcommand{\footnoterule}{} 
\caption{Photometric redshifts of PHOTO-z and LePHARE in the CFHTLS \texttt{D1}
field are compared to the spectroscopic sample consisting of VVDS
spectroscopic data (there are no SDSS or DEEP2 spectra). In this
table, $\sigma$ denotes the `true' dispersion, without clipping
outliers, and $\sigma_{\rm \Delta z/(1+z)}$ denotes the width of the
distributions after clipping outliers (as defined in
equations~\ref{sigclip} and introduced by Ilbert et al. 2006).  The
photometric redshifts of PHOTO-z where obtained with CFHTLS data only,
the LePHARE redshifts also make use of the VVDS BVRI imaging data and
(for 13 percent of objects) also of NIR data.
}
\label{tab:comp_LePHARE}
\centering
\renewcommand{\footnoterule}{}  
\begin{tabular}{llllllll}
\hline 
Code & Sample: CFHTLS-D1 & $N_{zspec}$  \footnote{Number of objects in the photometric redshift sample with spectroscopic redshifts}
                                                                                        &        Median error &        Mean error     &  $\sigma$ &  $\sigma_{\rm \Delta z/(1+z)}$   & $\eta
 [\%]$ \\
\hline
PHOTO-z
             &	all PHOTO-z objects 
                                                                          &	3035	&	-0.011	 &     -0.006   &	0.138	&      	0.033 &	 4.6$\%$   \\
PHOTO-z	     &	good PHOTO-z objects 
                                                                          &	2477	&       -0.010   &     -0.006   &      0.082	&      	0.031 &  1.8$\%$\\
PHOTO-z	     &	common good objects 
                                                                          &	2468	&       -0.010   &     -0.006   &      0.082	&      	0.031 &  1.7$\%$\\	
\hline                                                                                                                                                 
LePHARE    
            &	all LePHARE objects    
                                                                 	  &     3035	&	-0.003   &      0.015    &      0.187	&      	0.028 &  4.3$\%$	\\	
LePHARE     &	good LePHARE objects 
                                                  			  &     3012	&	-0.003	 &      0.013    &      0.173	&      	0.028 &   3.9$\%$	\\	
LePHARE     &	common good objects 
                                                                          &	2468	&	-0.003	 &      0.004   &      0.099 &      	0.029 &   2.6$\%$\\	
\hline                                                                                                                                                 
\end{tabular}

\begin{flushleft}
The definitions of the samples are: \\
	all PHOTO-z objects:    $z_{\rm spec}>0,z_{\rm PHOTO-z}>0$  \hfill\eject 
        good PHOTO-z objects:   $z_{\rm spec}>0,z_{\rm PHOTO-z}>0$, SED-type filtering, photometric redshift error filtering $\Delta z _{\rm photoz}< 0.25~(1+z_{\rm photoz})$ \\
        common good objects:    $z_{\rm spec}>0,z_{\rm PHOTO-z}>0$, SED-type filtering, photometric redshift error filtering for both codes \\
        all LePHARE objects:    $z_{\rm spec}>0,z_{\rm LePHARE}>0$ \\
        good LePHARE objects:   $z_{\rm spec}>0,z_{\rm LePHARE}>0 $, photometric redshift error filtering $ (\rm  (z_{\rm sup-{68}}- z_{\rm inf_{68}}) ) \le 0.25~(1+z_{phot}) $ \\
\end{flushleft}
\label{allz_table_D1}
\end{table*}
Table~\ref{allz_table_D1} and Fig.~\ref{FigLePHARE_PHOTO-z} 
show, that if one takes all spectroscopic
objects in the merged sample of the PHOTO-z and LePHARE catalogs,
the clipped dispersions and the outlier rates are similar
($\sigma_{\rm \Delta z/(1+z)}=0.033$, $\eta=4.6\%$ and $\sigma_{\rm
clip}=0.028$, $\eta=4.3\%$ for the PHOTO-z and LePHARE code
respectively), whereas the true dispersion of PHOTO-z is significantly
smaller than that of the LePHARE catalog ($\sigma=0.138$ vs $\sigma=0.187$).  
It also shows, that using SED-filtering and photometric redshift error
filtering (for the PHOTO-z case) and photometric redshift error
filtering (for the LePHARE case --note, that we don't make use of the
full probability function but just of the redshift range, including
$68\%$ of the redshift probability when defining `good objects' as $
(\rm (z_{\rm sup-{68}}- z_{\rm inf_{68}}) \le 0.25~(1+z_{phot}) $ for
the LePHARE catalog) reduces the outlier rate for both cases. The
stronger decrease of outliers in the PHOTO-z case ($1.8\%$ vs $3.9\%$)
is also caused by the effect that more objects are filtered out when
defining a `good object catalog', relative to the LePHARE catalog,
as can be seen from the sample size. To see how photometric redshifts
compare for objects which are considered as `good' objects in both
catalogs, we define a catalog of `common good objects' and
compare their photometric redshift quality in
Table~\ref{allz_table_D1} as well. For common good objects, the
outlier rate and the true dispersion is smaller using PHOTO-z
redshift, whereas the clipped sigma is slightly smaller for LePHARE
photometric redshifts. 
However the PHOTO-z redshifts show a small systematic effect compared
to the LePHARE ones, as the redshifts seem to slightly oscillate around the 
45 degree line. Whether this effect might be cured by optimizing the SED-templates 
for the CFHT filters will be investigated in an upcoming paper (Lerchster et al. in prep.), 
but one should keep in mind that the LePHARE redshifts are more immune against these 
oscillations since they do not use only the CFHT-$u^*g'r'i'z'$ data but also
include the CFH12K $BVRI$ data and in addition partly NIR data (J and K).
If one derives photometric redshifts for large
area surveys, one will not have a spectroscopic sample to compare
with; however comparing the redshift results and the assignemnts of
`good or secure photometric redshift objects' will improve the
selection of a robust sample with few outliers.
%
\section{Photometric Redshifts in the CFHTLS "Wide Fields"
\texttt{W1}, \texttt{W3} and \texttt{W4}}
\begin{figure} \centering
\includegraphics[width=6.cm]{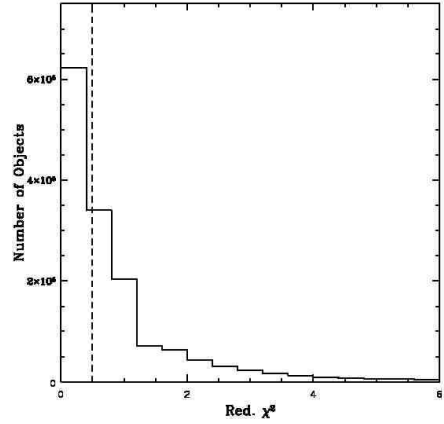}
\caption{Histogram of the reduced $\chi^2$ for all
galaxies in the CFHTLS Wide field \texttt{W1} as obtained for the best
fitting template and redshift. The dotted vertical line indicates the
median reduced $\chi^2$ of 0.5. }
\label{FigHistoChiError}
\end{figure} 

%
\begin{figure} \centering
\includegraphics[width=6.cm]{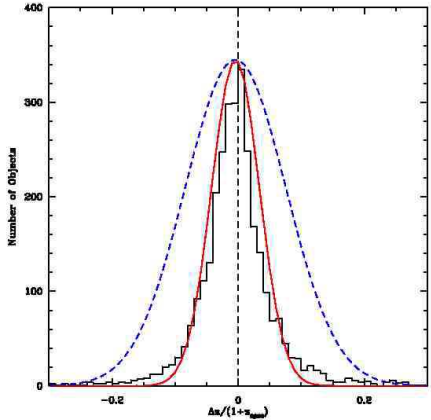}
\caption{Histogram of the photometric redshift errors in the CFHTLS
Wide field \texttt{W1}. The median redshift error is -0.004. The
values for the formal dispersion and the clipped dispersion are
$\sigma=0.08$ and $\sigma_{\rm \Delta z/(1+z)}=0.038$ for the W1 field. The
central distribution nearly is gaussian (with a width of $0.038$)
whereas the wings beyond $ |z_{\rm phot}-z_{\rm spec}|> 0.1$ cannot be
described with the same gaussian at all. To illustrate that, we have
added two gaussians with width of $0.038$ (in red) and $0.08$ (dashed
blue) and amplitudes matching the true error distribution at zero. The
high formal dispersion of $\sigma=0.08$ comes from the outliers; there
are 27 objects with $|z_{\rm phot}-z_{\rm spec}|> 0.3$ and 16 objects
with $|z_{\rm phot}-z_{\rm spec}|> 0.4$.  Considering both
$\sigma_{\rm \Delta z/(1+z)}$ and $\sigma $ tells, how `severe' the outliers are.}
\label{FigW1zError}
\end{figure} 
%
%
%

We obtain photometric redshifts after zeropoint calibration as
described before.  A comparison between spectroscopic (VVDS sample
only) and photometric redshifts for the CHTLS Wide Field \texttt{W1}
and \texttt{W4} is shown in Fig.~\ref{FigZZ}.  The outlier fraction is
$\eta \leq 4\%$.  In Fig.~\ref{FigW1zError} we show 
 the distribution of the redshift errors on CFHTLS Wide field
\texttt{W1} using the VVDS spectra. The error distribution is only
gaussian in its center, which has a width of $0.038$ equal to the
`clipped width', $\sigma_{\rm \Delta z/(1+z)}$ defined by Ilbert et al. \cite{ilbert06}. The outliers
are too many to be compatible with such a narrow gaussian; the true
dispersion is equal to $0.08$ (for galaxies with VVDS spectra in the
\texttt{W1}-field)

Fig.~\ref{FigHistoChiError} presents the $\chi^2$
distribution of the best fitting templates and photometric redshifts
for all the objects. The median value of the reduced $\chi^2$ is 0.5
which implies that the galaxy templates describe the galaxies rather
well.
\subsection{Tests of the zeropoint calibration method} 
\begin{figure}
\centering
\includegraphics[width=8.5cm]{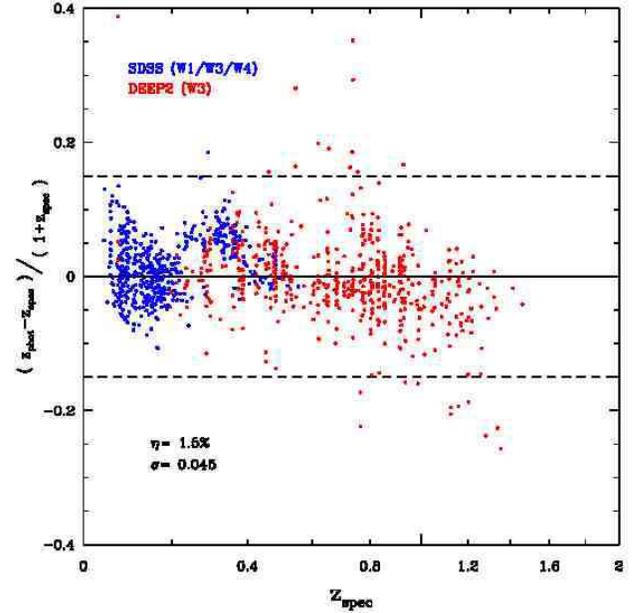}  
\caption{Photometric redshifts in the CFHTLS Wide fields against the
spectroscopic redshifts of the SDSS DR6 (blue dots) and DEEP2 (red
dots).  The dotted lines are for $z_{phot}=z_{spec}\pm
0.15~(1+z_{spec})$. These spectroscopic data have not been used for
the calibration of zeropoint shifts, and thus provide an independent
estimate of the redshift accuray.}
\label{FigCompSDSS}
\end{figure}

We now investigate how well our empirical calibration of zeropoints
using the color-color diagram of stars works. We derive zeropoints
offset from matching the color-color diagram of stars as learned in
the \texttt{W1} and \texttt{W4} subfields fields with VIMOS spectroscopic data. This
offsets are used in the photoz code.  We then compare our photometric
redshifts to 410 galaxies from the Deep survey (Weiner et
al. 2005; Davis et al. 2007) in the CFHTLS
\texttt{W3} field (which cover areas different from the VIMOS
galaxies).  The photometric redshift prediction for these Deep
galaxies is accurate to $\sigma_{\rm \Delta z/(1+z)} \sim 0.041$ with an outlier
rate of $\eta \sim 6\%$.
\\ 
Since our data (partly) overlap with the Sloan Digital Sky
Survey (SDSS), we can compare the photozs to spectroscopic ones for
further 528 galaxies from the SDSS DR6. These low redshift objects give
an accuracy of $\sigma_{\rm \Delta z/(1+z)} \sim 0.036 $ and a outlier rate of
only $1/528$. When we finally combine the DEEP2 and the SDSS DR6
spectra we determine an accuracy of $\sigma_{\rm \Delta z/(1+z)} \sim 0.045$ 
and an outlier rate of $\eta \sim 1.5\%$ for the combined sample, see
Fig. \ref{FigCompSDSS}.  
\\ 
We think, that this shows, that our
photometric calibration method can be applied to all CFHTLS Wide
fields.
\\
\begin{figure*}
\centering
\includegraphics[width=8.5cm]{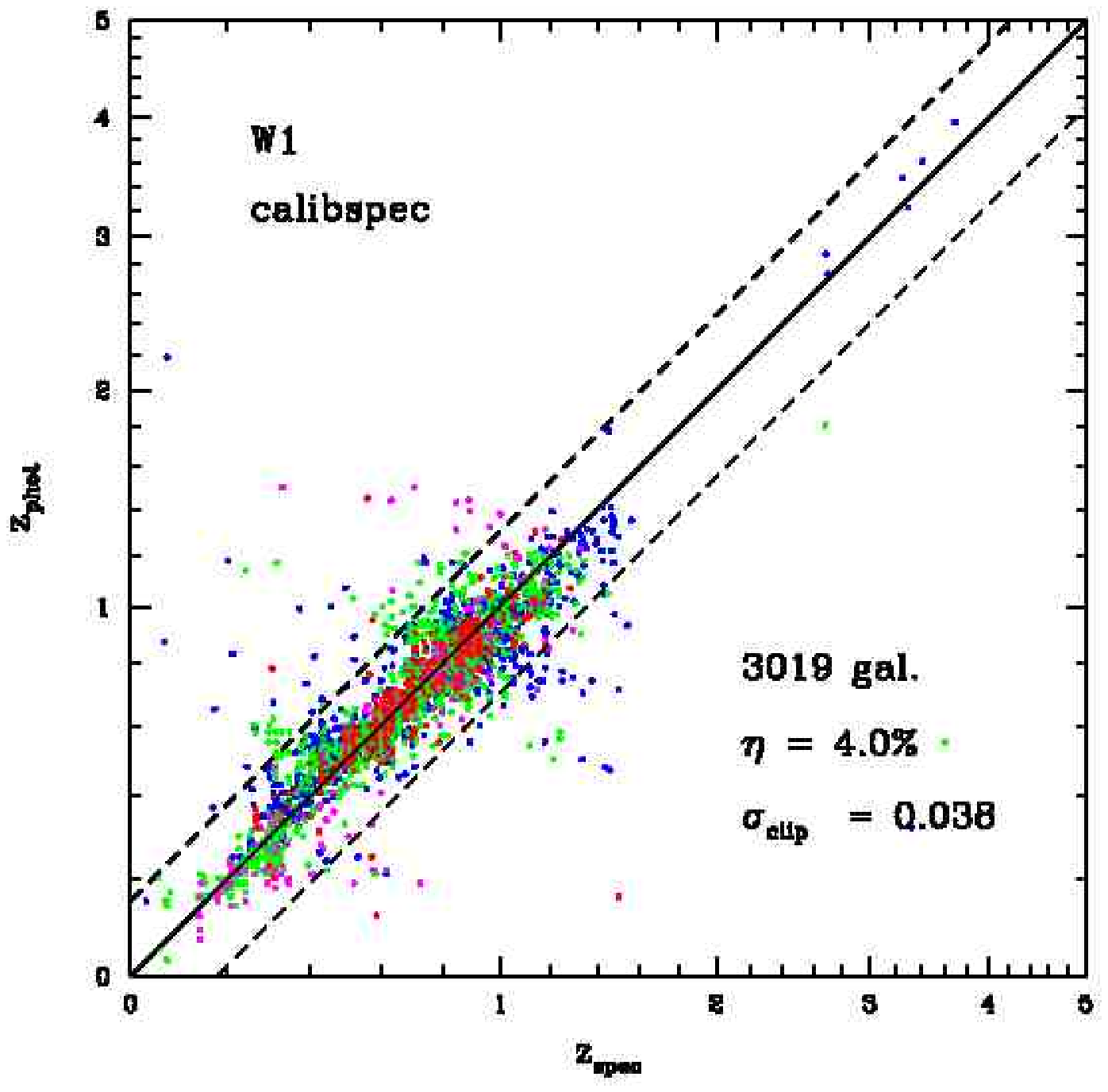}
\includegraphics[width=8.5cm]{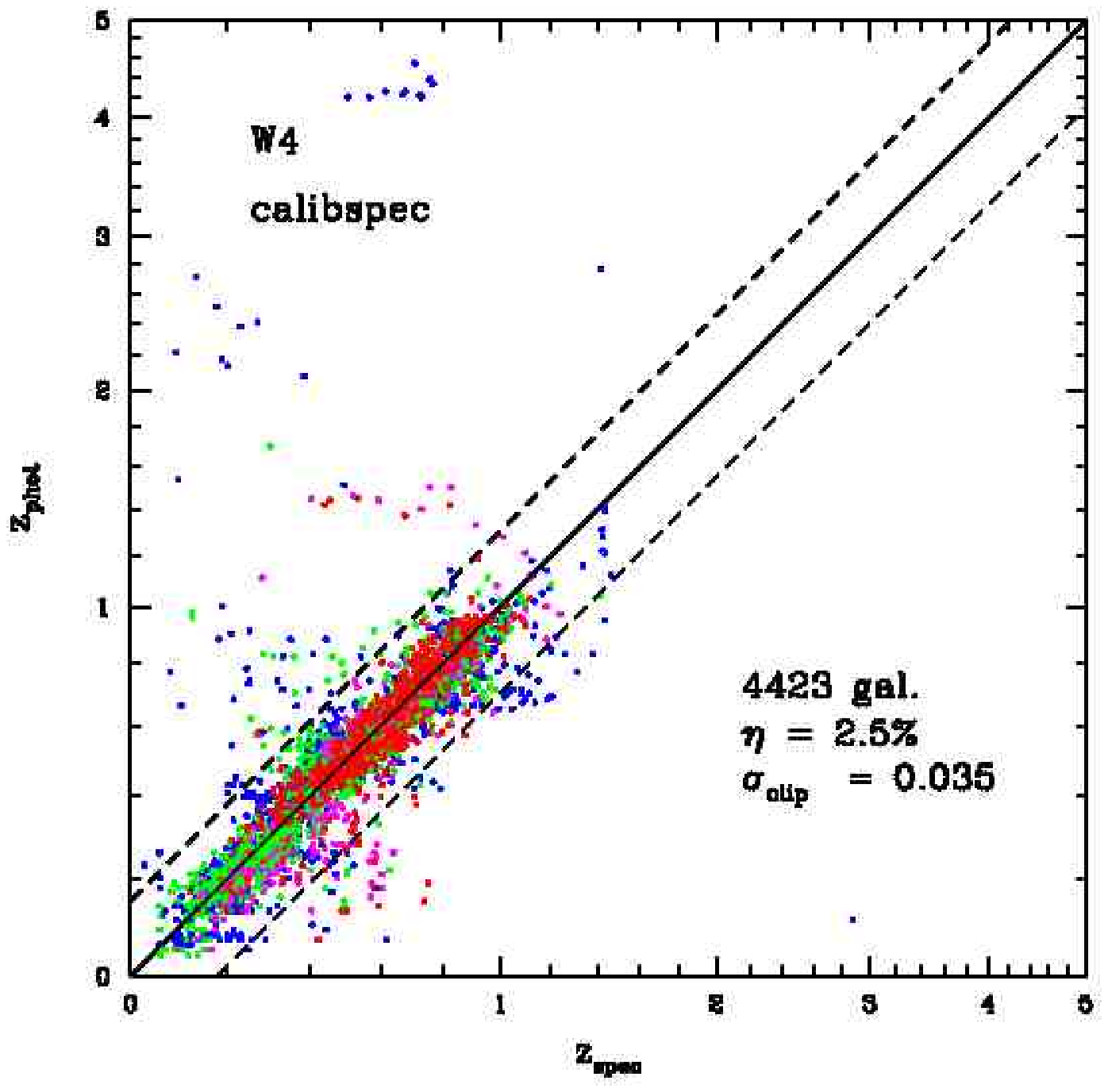}  
\caption{PHOTO-z photometric redshifts against spectroscopic ones for
  the CFHTLS \texttt{W1}(left) and \texttt{W4} (right) field.  SED
  types are color-coded: red for E/S0 galaxy types, blue for strongly
  star forming, blue galaxies, magenta and green symbols are for SEDs that
  form  a continuous sequence of SEDs in color and star forming activity 
  between the red and blue galaxies. The green symbols are for SEDs bluer
  than those coded with magenta symbols.
  The dotted lines are for $z_{phot}=z_{spec}\pm
  0.15 ~(1+z_{spec})$. The fraction $\eta$ of catastrophic outliers is
  defined as as fraction of galaxies for which
  $|z_{spec}-z_{phot}|/(1+z_{spec}) > 0.15$ holds.  $\sigma_{\rm \Delta z/(1+z)}$ is a
  measure for the redshift accuracy, defined as $\sigma_{\rm \Delta z/(1+z)}=1.48~ \times
  ~median (|z_{\rm spec}-z_{\rm zpec}|/(1+z_{\rm spec}))$, applied to non-outliers only.  This
  follows the definition of Ilbert et al. \cite{ilbert06}.  This `variance'
  $\sigma_{\rm \Delta z/(1+z)}$ equals $ 0.038$ for the \texttt{W1} and 0.035 for the
  \texttt{W4} field.
For this example we have been using the spectroscopic data to 
calibrate the zeropoint offsets.
}
\label{FigZZ}
\end{figure*}
%
%
%
\begin{figure*}
\centering
\includegraphics[width=8.5cm]{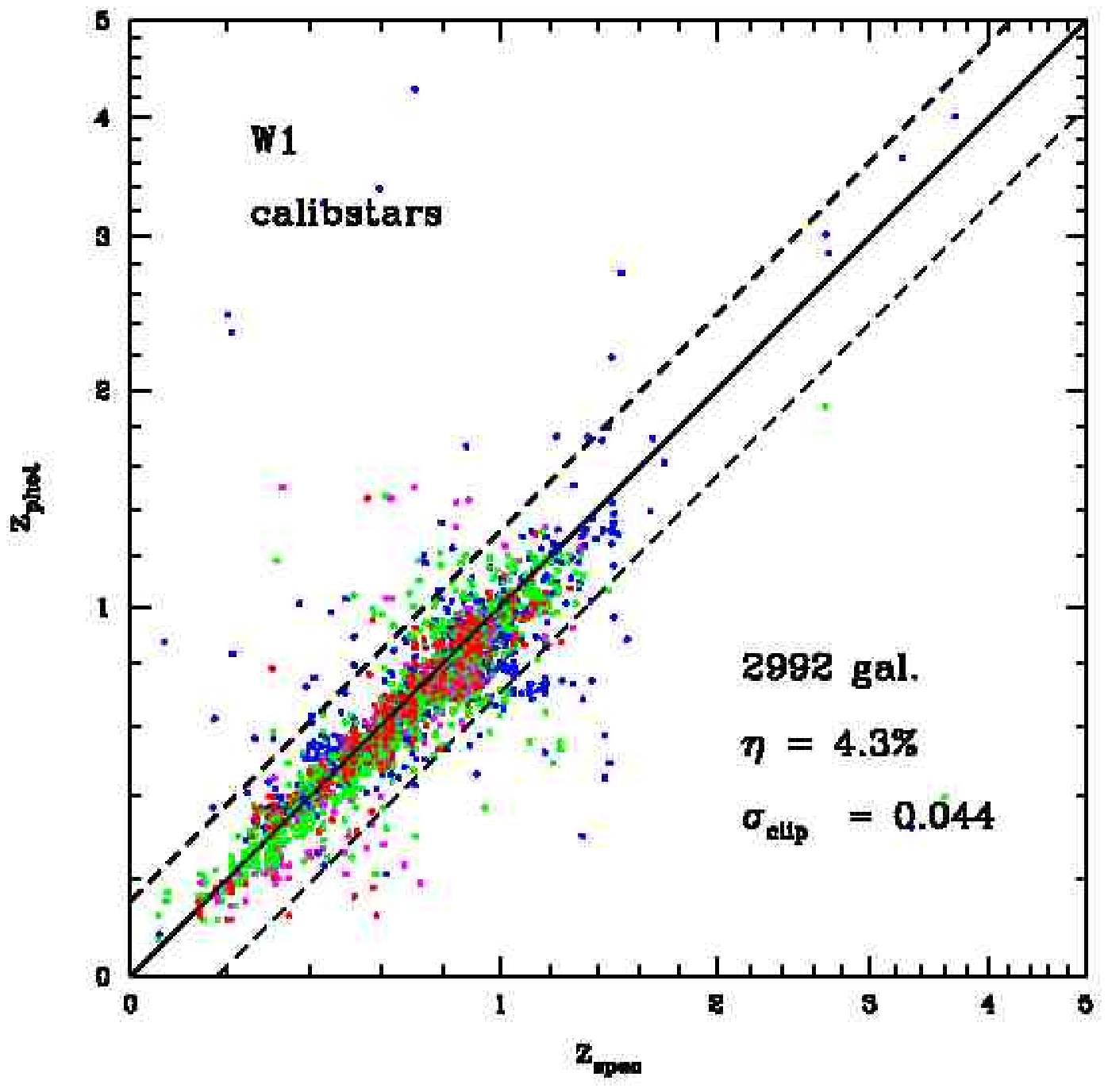}
\includegraphics[width=8.5cm]{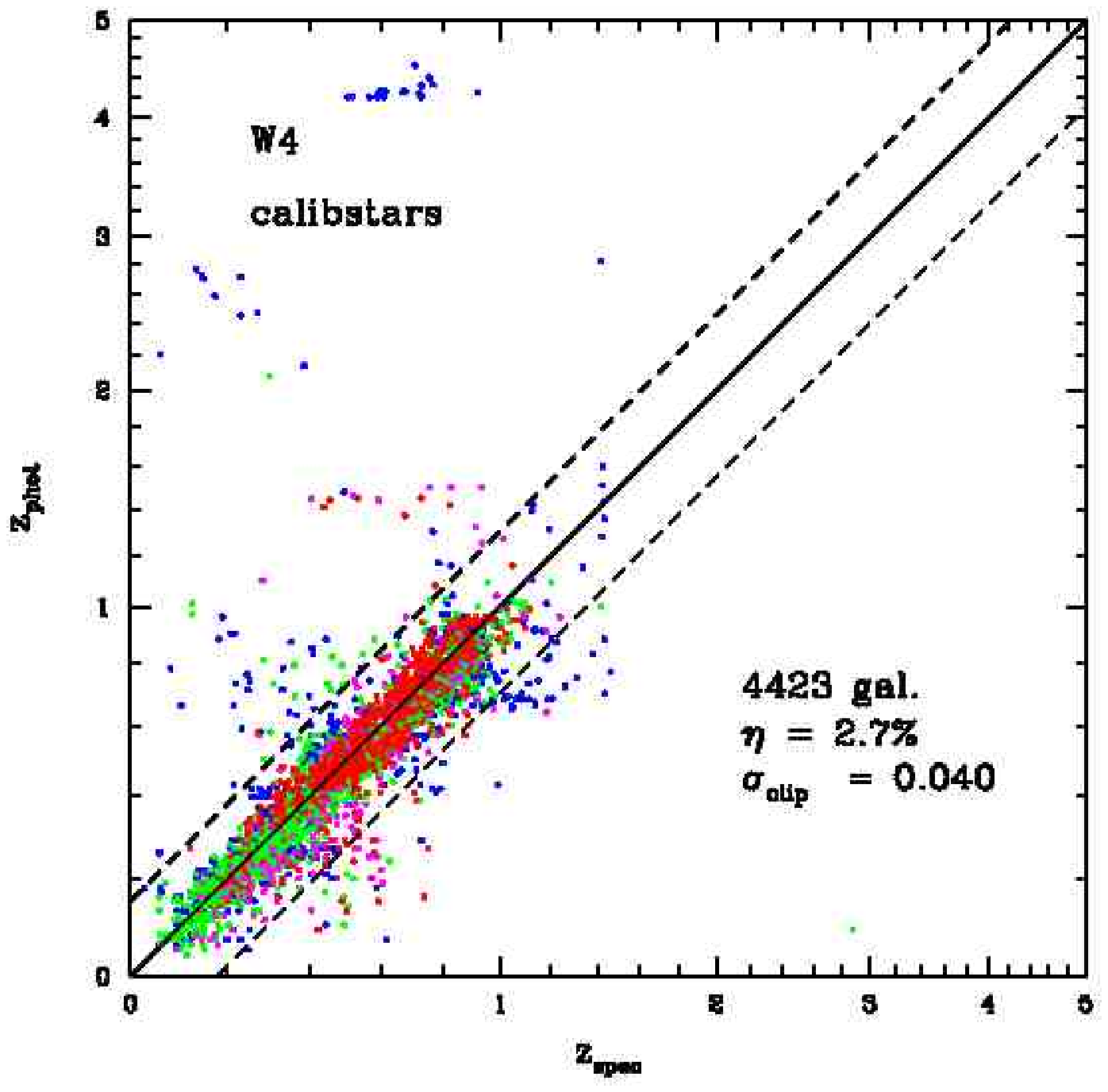}   
\caption{Same as Fig.~\ref{FigZZ}, this time we use the stellar colors
  and no spectroscopic information for the zeropoint calibration. The
  results become slightly worse/ Left panel: result for the CFHTLS
  \texttt{W1} field. Right panel: result for the CFHTLS \texttt{W4}
  field. Color coding for SED-types is as above.}
\label{FigPhotSpecCompW1}
\end{figure*}    
We show in Fig.~\ref{FigPhotSpecCompW1} how the photometric redshifts
change if we use for the zeropoint calibrations stars only (instead of
galaxy redshifts, see Fig.~\ref{FigZZ} for that case). Results are
shown for the CFHTLS \texttt{W1} field and the CFHTLS \texttt{W4}
field. The accuracy deteriorates from $\sigma_{\rm \Delta z/(1+z)}=0.035$
($\sigma_{\rm \Delta z/(1+z)}=0.038$) to $\sigma_{\rm \Delta z/(1+z)}=0.04$ ($\sigma_{\rm
clip}=0.044$) for the field \texttt{W4} (\texttt{W1}), the outlier rate roughly stays
the same.

%
%
\subsection{Redshift accuracies as a function of type \& brightness}
\begin{figure*}
\centering
\includegraphics[width=11.5cm]{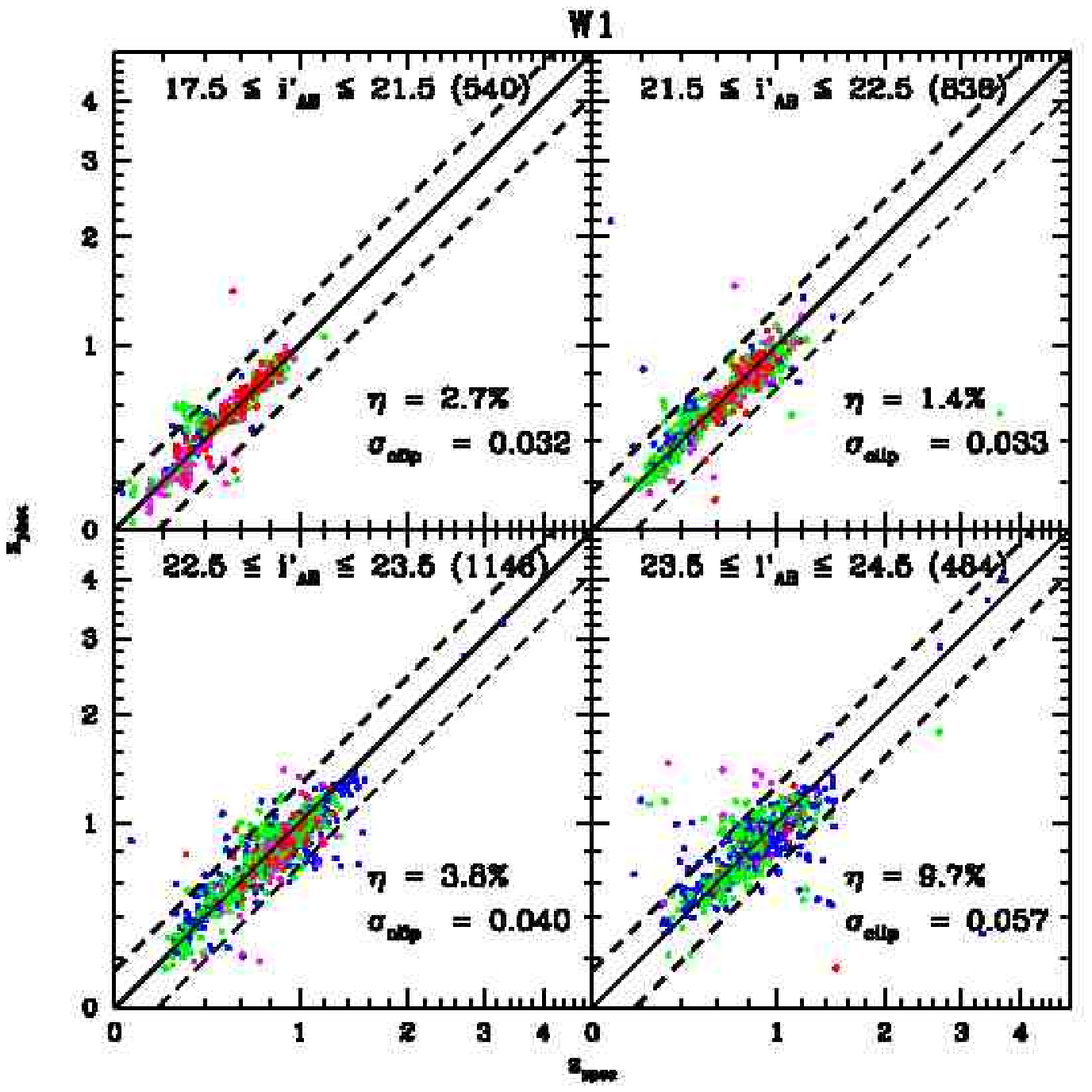} 
\includegraphics[width=11.5cm]{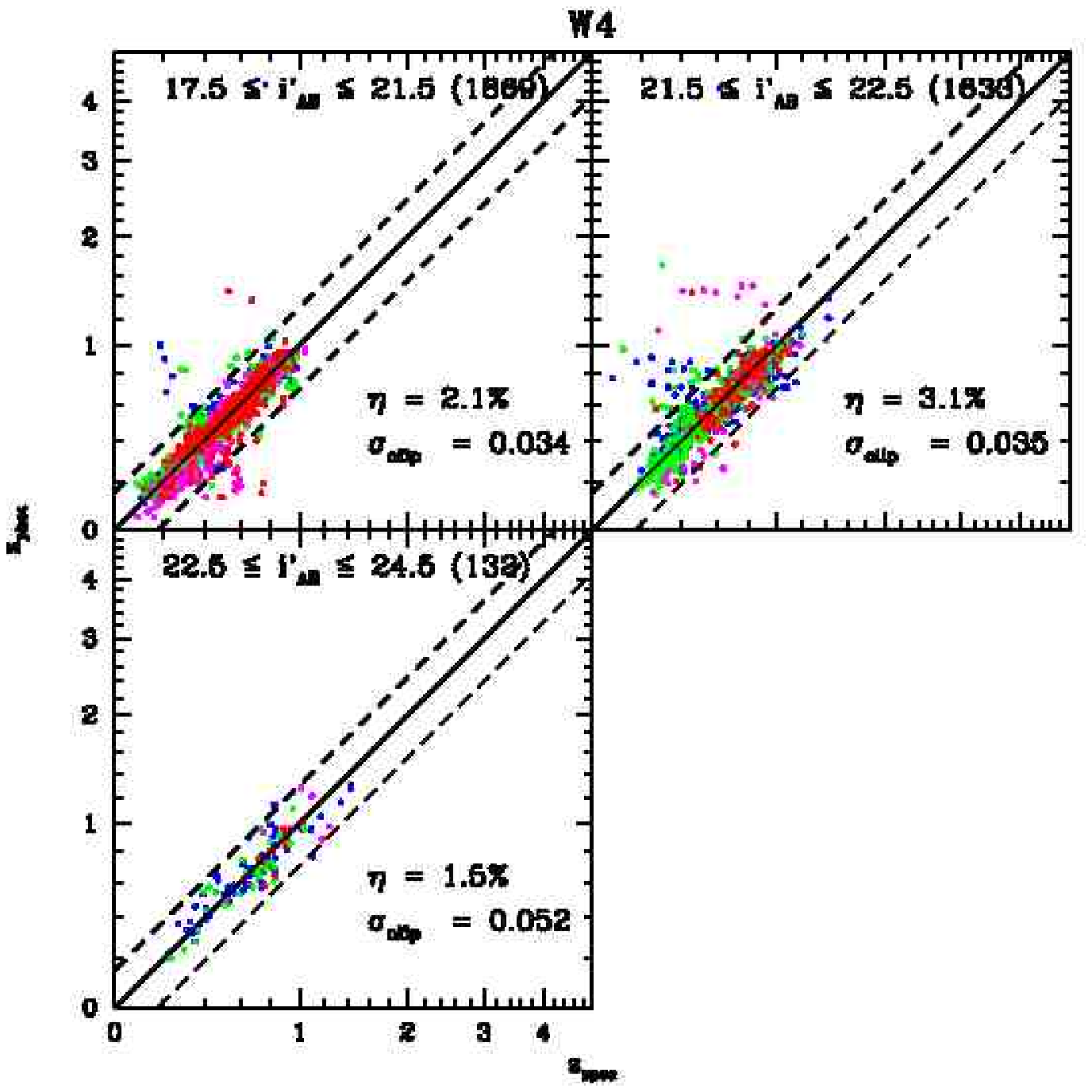}     
\caption{Same as Fig.~\ref{FigZZ} (i.e., calibration on spectra). Each panel shows subsamples with different magnitudes. At the top the panels for the CFHTLS \texttt{W1} field and at the bottom the CFHTLS \texttt{W4} field are shown.} 
\label{FigPhotSpecCompMag}
\end{figure*}

\begin{figure*}
\centering
\includegraphics[width=11.5cm]{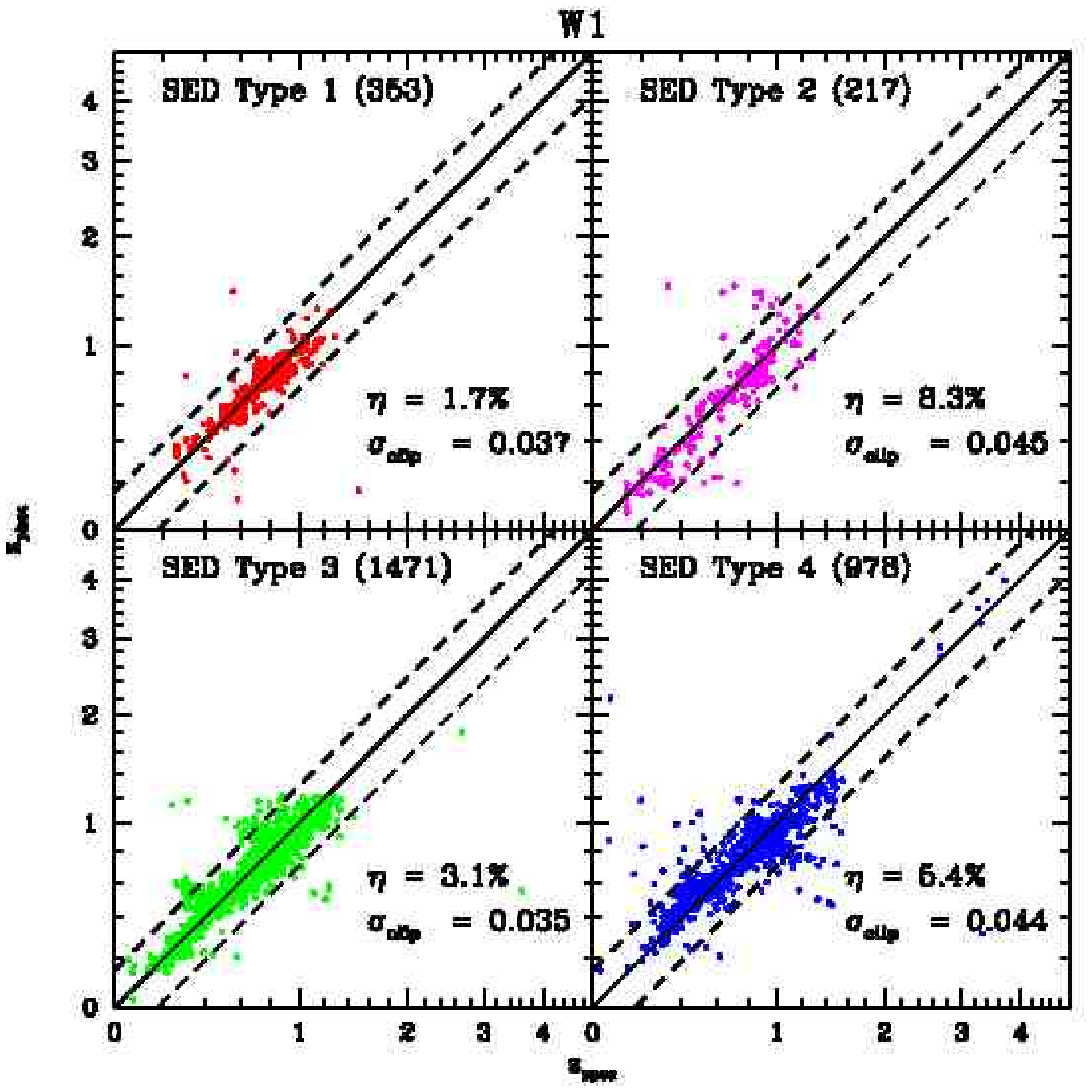} 
\includegraphics[width=11.5cm]{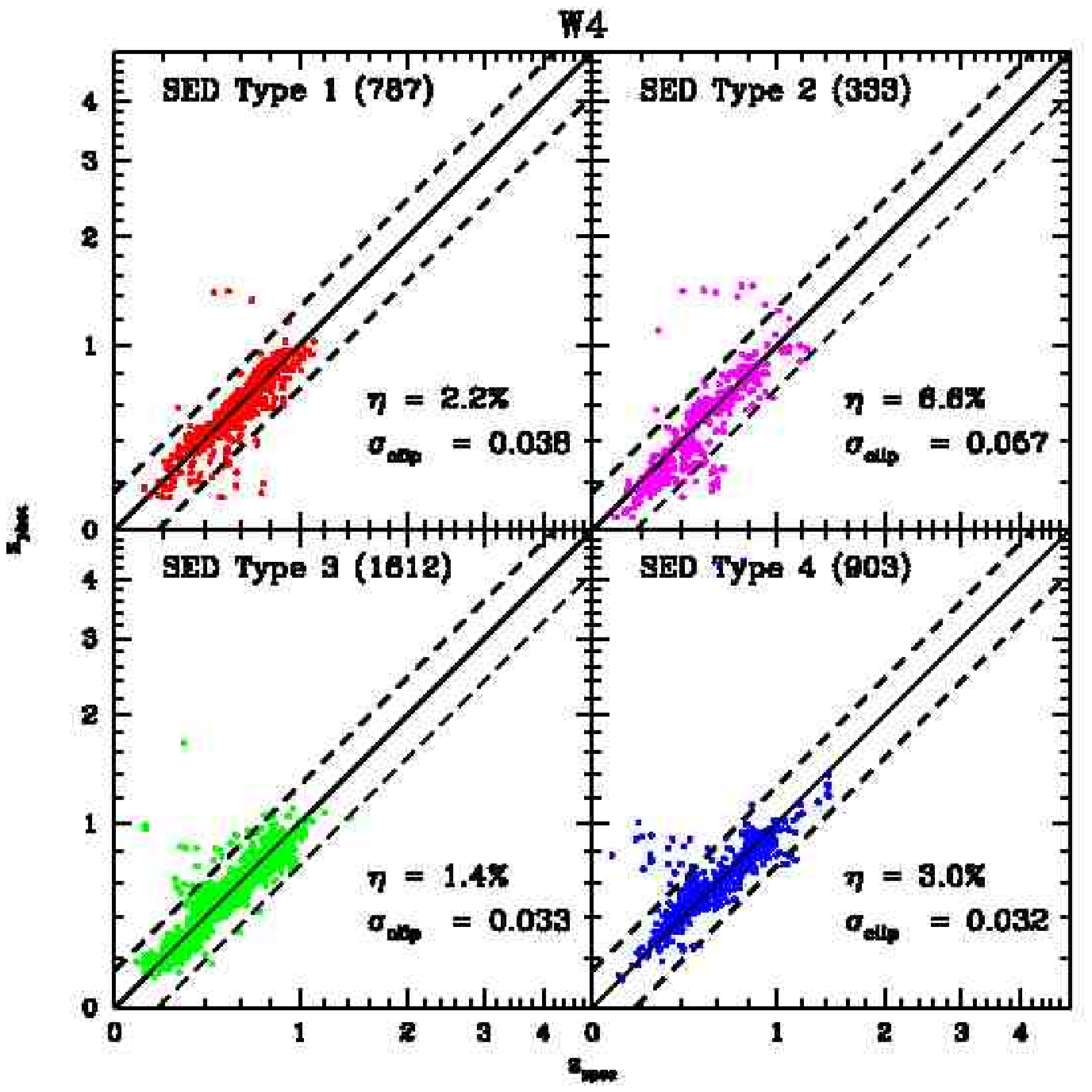}   
\caption{Same as Fig.~\ref{FigZZ} (i.e., calibration on spectra).
At the top panels the CFHTLS \texttt{W1} field, selected at $17.5 \leq i_{AB} \leq 24$ 
and at the bottom panels the CFHTLS \texttt{W4} field, selected at $17.5 \leq i_{AB} \leq 22.5$ 
is shown. Each panel shows a subgroup of 
galaxies which were sorted into four SED-groups 
according to the best-fitting template:
The first group (red symbols)  contains SEDs that describe 
ellipticals and S0s and is
the fourth group (blue symbols) contains very blue, strongly starforming
SEDs  and the second (magenta symbols) and third (green symbols) 
group form  a continuous sequence of SEDs in color and star forming activity 
 between the first and fourth group.}
\label{FigPhotSpecCompSED}
\end{figure*}

In Fig.~\ref{FigPhotSpecCompMag} we compare the photometric redshifts
to the spectroscopic redshifts for different apparent magnitude
intervals for the CFHTLS \texttt{W1} and \texttt{W4} fields.  The
fraction of catastrophic outliers $\eta$ increases from 2.7 \% to 9.7
\%, going from $17.5 \leq i'_{AB} \leq 21.5$ up to $23.5 \leq i'_{AB}
\leq 24.5$, this has been also seen by Ilbert et al. \cite{ilbert06}.
\\ 
Fig.~\ref{FigPhotSpecCompSED} shows the photometric redshifts
versus the spectroscopic redshifts for different spectral types for
the CFHTLS \texttt{W1} and \texttt{W4} field.  We sort the SEDs we use
to describe galaxies into 4 groups.  The first one contains SEDs that
describe ellipticals and S0s and is colored red in plots, the fourth
contains very blue, strongly starforming SEDs (colored blue in plots)
and the third (magenta) and fourth (green) group form a continuous
sequence of SEDs in color and star forming activity between the first
and fourth group of galaxy-SEDs.  The accuracies of the photometric
redshifts become $\sigma_{\rm \Delta z/(1+z)} \sim 0.037$ for group one and two
(red and magenta), and $\sigma_{\rm \Delta z/(1+z)} \sim 0.044$ for group three
and four (green and blue).  The catastrophic outliers increase by a
factor of about three from group old (old SEDs) to the other groups
(younger SEDs) which has also been found by Ilbert et
al. \cite{ilbert06}.
\\
It is worth to note, that the integration time eg. of the i'-band on
the Deep Field \texttt{D1} is $>$ 35 times longer then integration
time of the i'-band data in the Wide Field \texttt{W1}. Nevertheless, the
outlier rate is only slightly larger 
 and the accuracy is almost the same in the shallower sample
(if one limits the comparison sample to $I<22.5$).  In other words: in
order to obtain photometric redshifts for $I<22.5$ galaxies, it does not
play a role whether the data are `rather deep' (\texttt{W1} , with $I<24.5$) or
`very deep' (\texttt{D1} -T0003, with $I<25.9$). For these depths the photon noise is
not relevant any longer, but solely how well the templates can
reproduce the true galaxy SEDs.
%
\begin{table}
\begin{minipage}[t][]{\columnwidth}
\caption{Photometric redshift accuracy seperated for each subfield (this
makes use of the VVDS spectra only)}
\label{tab:subfieldcheck}
\centering
\renewcommand{\footnoterule}{}  
\begin{tabular}{lll}
\hline
Field    &  $\sigma_{\rm \Delta z/(1+z)}$   & $\eta$ [\%] \\
\hline
\hline
\texttt{W1p2p2} &  0.042 & 4.0 \\
\texttt{W1p2p3} &  0.038 & 4.0 \\
\texttt{W4m0m0} &  0.034 & 1.3 \\
\texttt{W4m0m1} &  0.030 & 2.8 \\
\texttt{W4m0m2} &  0.033 & 1.9 \\
\texttt{W4p1m0} &  0.034 & 0.7 \\
\texttt{W4p1m1} &  0.036 & 3.4 \\
\texttt{W4p1m2} &  0.043 & 2.1 \\
\texttt{W4p2m0} &  0.032 & 2.4 \\
\texttt{W4p2m1} &  0.037 & 3.0 \\
\texttt{W4p2m2} &  0.031 & 2.4 \\
\hline
\end{tabular}
\end{minipage}
\end{table}
\subsection{Redshift distributions}
\begin{figure}
\centering
\includegraphics[width=8.cm]{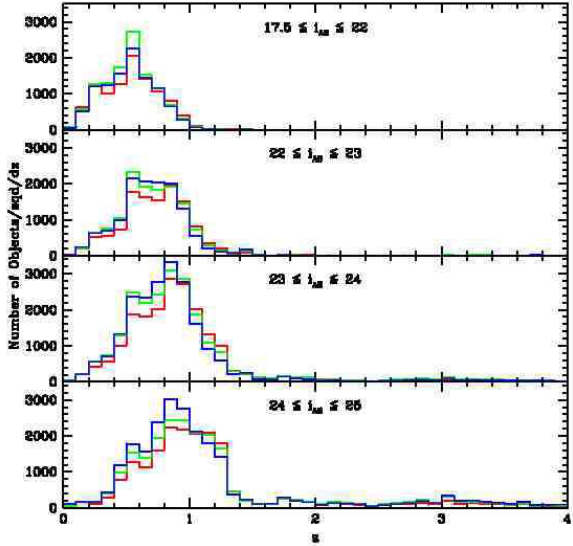}  
\caption{Photometric redshift distributions of `good' objects 
in the three CFHTLS Wide fields, \texttt{W1} (red line), \texttt{W3} (blue line) and \texttt{W4} (green line). The redshift distributions are shown from bright to faint selected samples.}
\label{FigHistoGalFields}
\end{figure}
\begin{figure}
\centering
\includegraphics[width=8.cm]{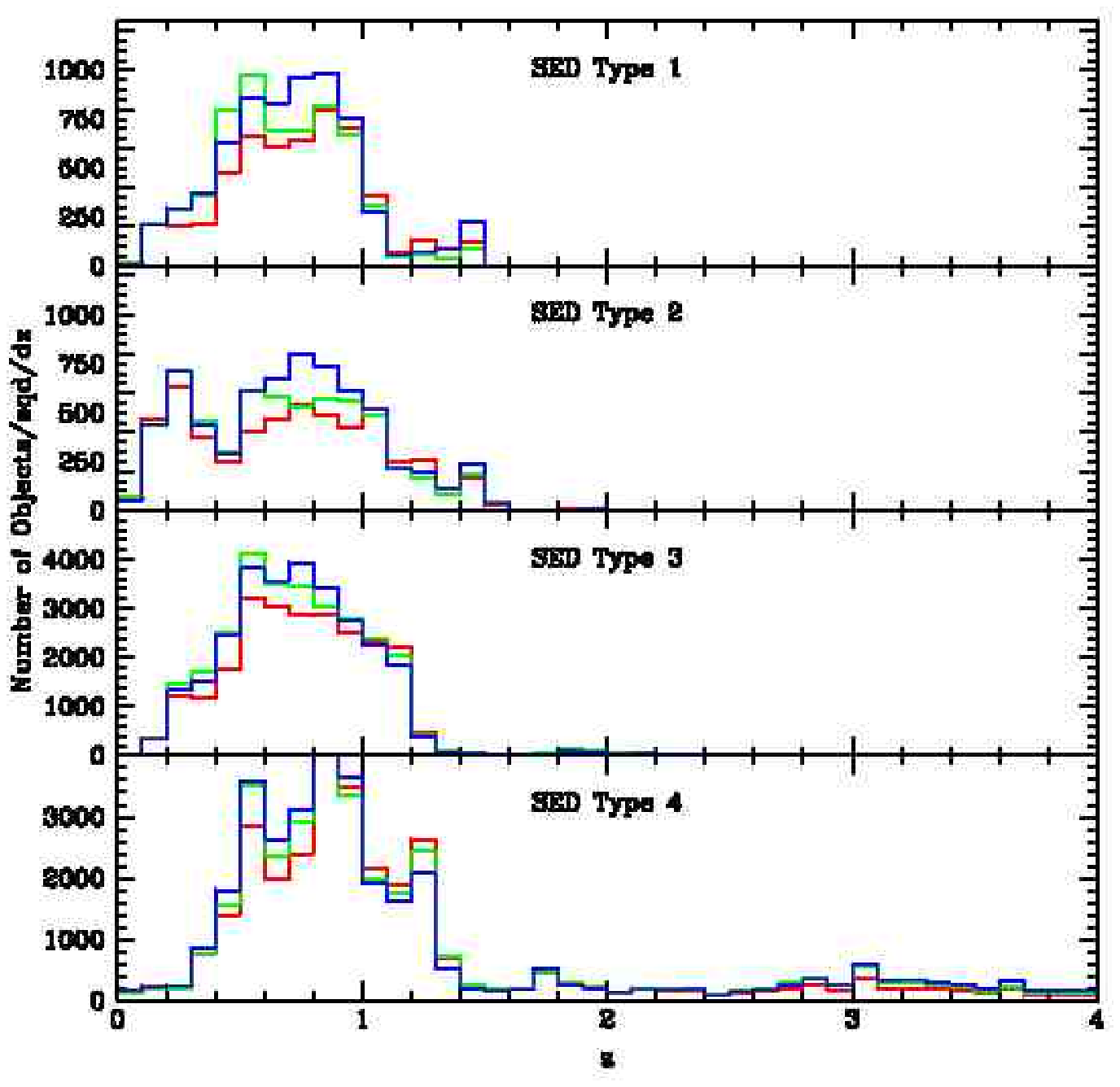}   
\caption{Same as Fig.~\ref{FigHistoGalFields} , 
but each panel shows a different selection of spectral 
types, according to the best-fitting template. From top to the bottom: 
The first group  contains SEDs that describe 
ellipticals and S0s and the fourth group contains very blue, 
strongly starforming
SEDs; the second  and third  
group form  a continuous sequence of SEDs in color and star forming activity 
between the first and fourth group.}
\label{FigHistoGalFieldsSED}
\end{figure}
Fig.~\ref{FigHistoGalFields} shows the galaxy redshift histogram of
all objects in the four CFHTLS Wide Fields. The median redshift (see
Tab.~\ref{tab:meanred}) is in good agreement in the four fields
although the redshift distribution in the \texttt{W1} field is shifted to
higher redshift. In Fig.~\ref{FigHistoGalFieldsSED} the SED redshift
distribution of all galaxies in the CFHTLS Wide Field \texttt{W1} and
\texttt{W4} is shown.
\begin{figure}
\centering
\includegraphics[width=8.cm]{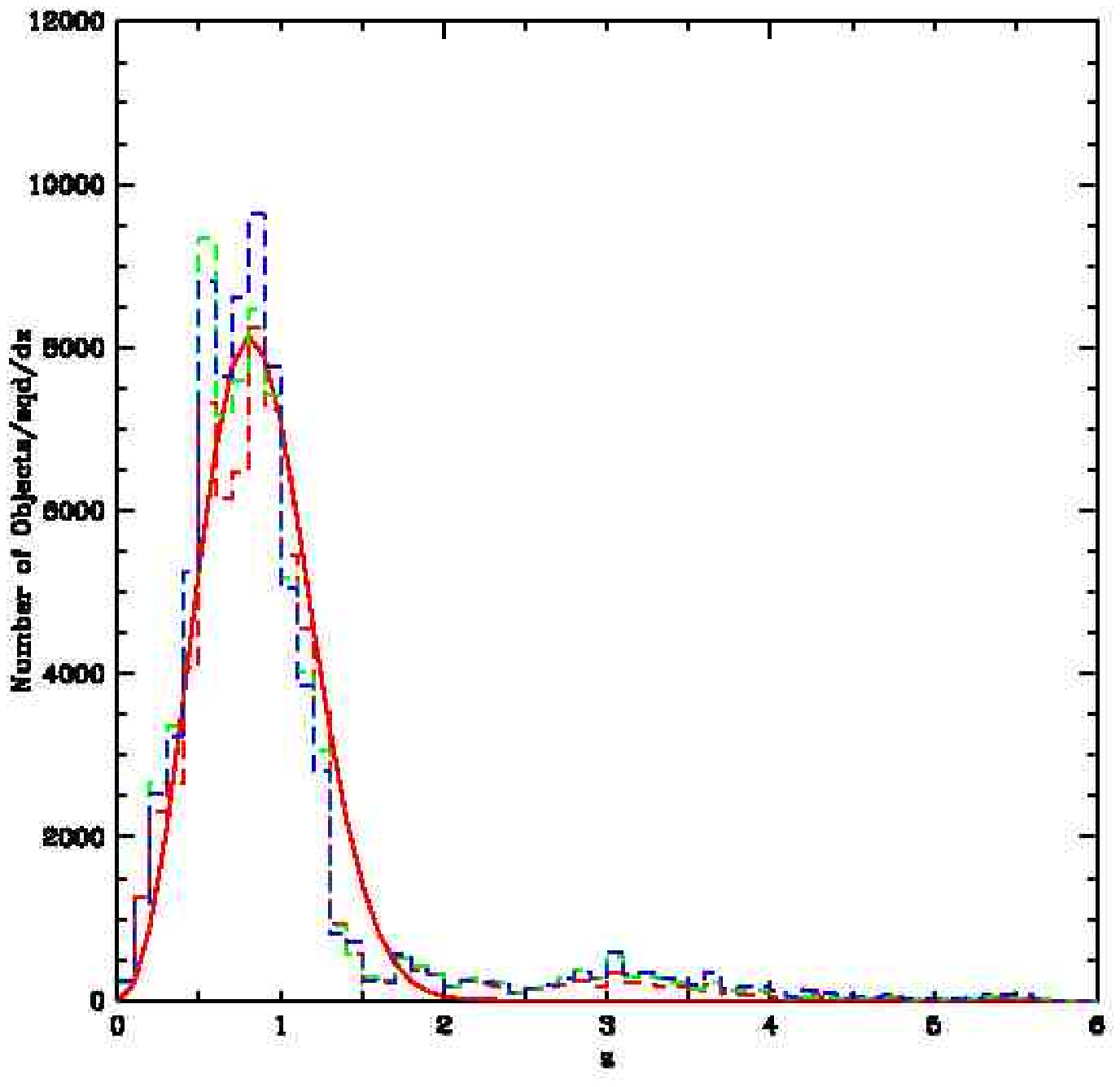}  
\caption{Redshift number distributions for galaxies in the three
CFHTLS Wide fields. The distributions are red, blue and green for the
\texttt{W1}, \texttt{W3} and \texttt{W4} fields, the median redshifts
in these fields are $z_{median, W1}=0.84$, $z_{median,W3}=0.79$ and
$z_{median, W4}=0.80$.  The red solid curve is a fit to the mean
galaxy distribution for the \texttt{W1}-field using the parametric
description given, e.g., in Van Warbeke et al. \cite{waerbeke01}.}
\label{FigHistoGal}
\end{figure} 
%
%
\begin{table}
\begin{minipage}[t][]{\columnwidth}
\caption{Median redshifts in the three CFHTLS Wide Fields (columns) for samples
 selected according to $17.5\leq i'_{AB} \leq 22$, $22\leq i'_{AB} \leq 23$, $23\leq i'_{AB} \leq 24$, and  $24\leq i'_{AB} \leq 25$ from the top to the bottom. Only galaxies with `good' photometric redshifts are considered.}
\label{tab:meanred}
\centering
\renewcommand{\footnoterule}{}  
\begin{tabular}{lccccc}
\hline \hline
Magnitude interval & $z_{\rm median}[\texttt{W1}]$    &  $z_{\rm median}[\texttt{W3}]$    & $z_{\rm median}[\texttt{W4}]$  \\
\hline
$17.5\leq i'_{AB} \leq 22.0$  & 0.53  &  0.53  &  0.54\\
$22.0\leq i'_{AB} \leq 23.0$  & 0.70  &  0.68  &  0.68\\
$23.0\leq i'_{AB} \leq 24.0$  & 0.77  &  0.75  &  0.74\\
$24.0\leq i'_{AB} \leq 25.0$  & 0.79  &  0.76  &  0.76\\   
\hline
\end{tabular}
\end{minipage}
\end{table}
\begin{table}
\begin{minipage}[t][]{\columnwidth}
\caption{Galaxy redshift distribution for objects with `good' photometric redshifts in the CFHTLS Wide fields, using the parametrisation of Van Waerbeke et al. \cite{waerbeke01}.}
\centering
\renewcommand{\footnoterule}{}  
\begin{tabular}{lccccc}
\hline \hline
Field & $z_0$     &  $\alpha$   & $\beta$ \\
\hline
\texttt{W1}  & 0.84  &  2.2   &  2.4\\
\texttt{W3}  & 0.79  &  2.2   &  2.4\\
\texttt{W4}  & 0.80  &  2.2   &  2.4\\
\hline
\end{tabular}
\end{minipage}
\label{tab:reddist}
\end{table}
%
In Fig.~\ref{FigHistoGal} the galaxy redshift histogram of all objects
in the CFHTLS Wide is shown. The galaxy redshift distribution can be
parameterized, following Van Waerbeke et al. \cite{waerbeke01}:
\begin{equation} n(z_s)=\frac{\beta}{z_0
\Gamma\bigl(\frac{1+\alpha}{\beta}\bigr)}\bigl(\frac{z_s}{z_0}\bigr)^{\alpha}
exp\bigl[-\bigl(\frac{z_s}{z_0}\bigr)^{\beta}\bigr],
\end{equation}
where $(z_0, \alpha, \beta)$ are free parameters. The best fitting
values for the 3 fields (\texttt{W1;W3;W4}) are show in Table
\ref{tab:reddist}. It can be seen from median redshift that the three
fields were observed to different depths.

%
%
\section{Comparing CFHTLS-W photozs from different methods  }
\begin{figure*}
\centering
\includegraphics[width=12.5cm]{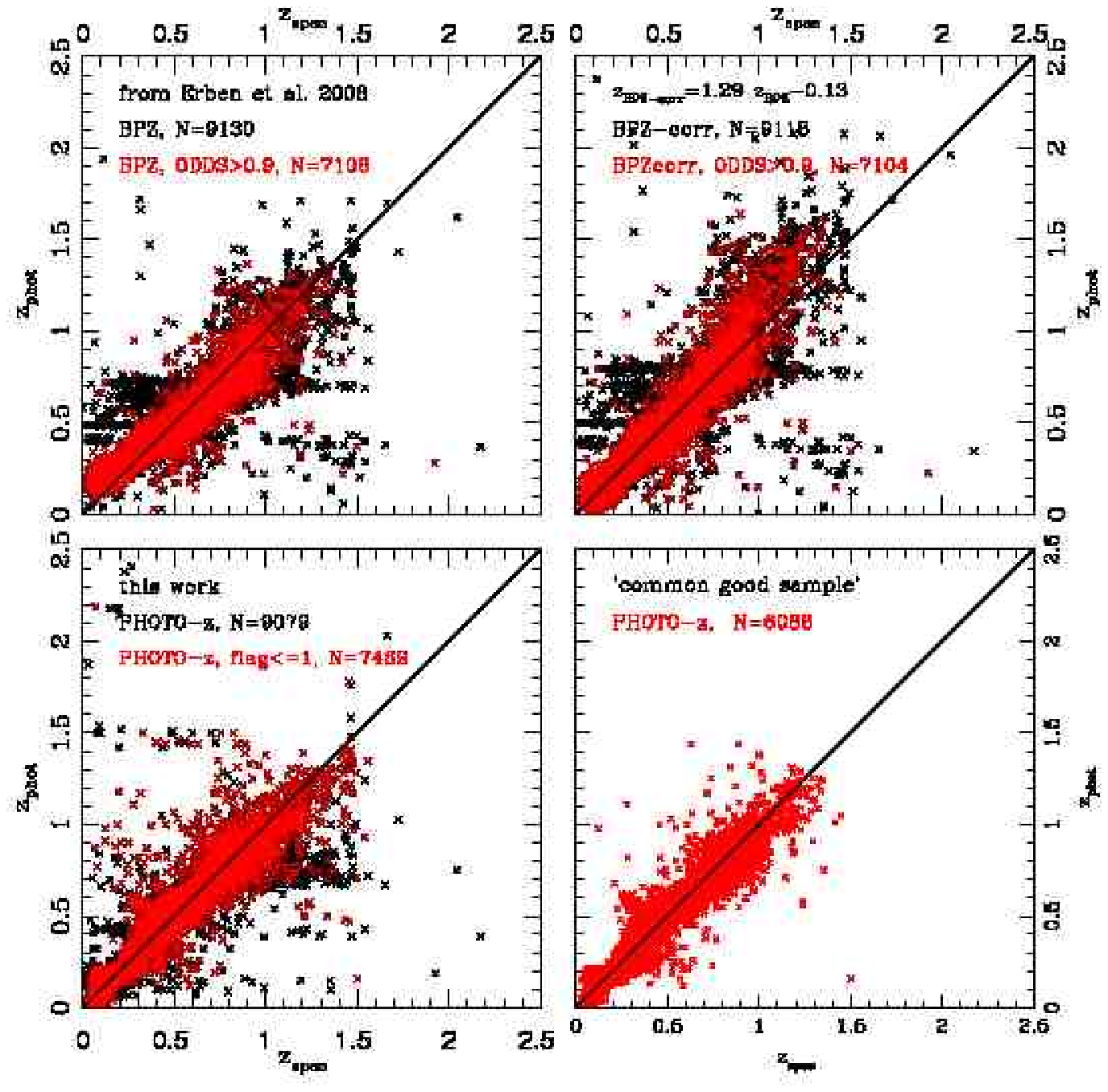}
\caption{This figure relates the spectroscopic and photometric
redshifts for the BPZ-method (upper left), the bias corrected
BPZ-redshifts (upper right) and the PHOTO-z redshifts of the Bender et
al. \cite{bender01} code (lower left). In each case, the full sample is shown with
black points, the sample which is considered as `good' is shown with
red points: these are in the case of the BPZ method all objects with
ODDS$>0.9$, and in the case of the PHOTO-z method all objects with
$\Delta z_{\rm phot}< 0.25*(1+z_{\rm phot})$ after excluding
starforming SEDs.  The BPZ redshifts show a clear bias as reported in
Erben et al. \cite{erben08} already. This bias is almost linear in
redshift, and can be compensated for $z<1.1$ with the equation given
in the upper right panel. For higher redshift, this compensation leads
to an overcorrection. The redshifts from the PHOTO-z code (lower left
panel) are free of bias. They have a few more outliers, which is
partly due to the fact, that there are more objects in this sample. At
redshifts above 1.2 there might be a very small bias towards
underestimating redshifts, there are, however, not many galaxies
left. The lower right panel finally shows the photometric redshifts
for the PHOTO-z method, were only objects that are considered as
`good' both in the BPZ, the BPZ-corrected and the PHOTO-z sample are
shown. The exact criteria for the sample selection can be read off
from Table~\ref{allz_table}.
   }
\label{specz_photoz_all_objects_1}
\end{figure*}
\begin{figure*}
\centering
\includegraphics[width=12.5cm]{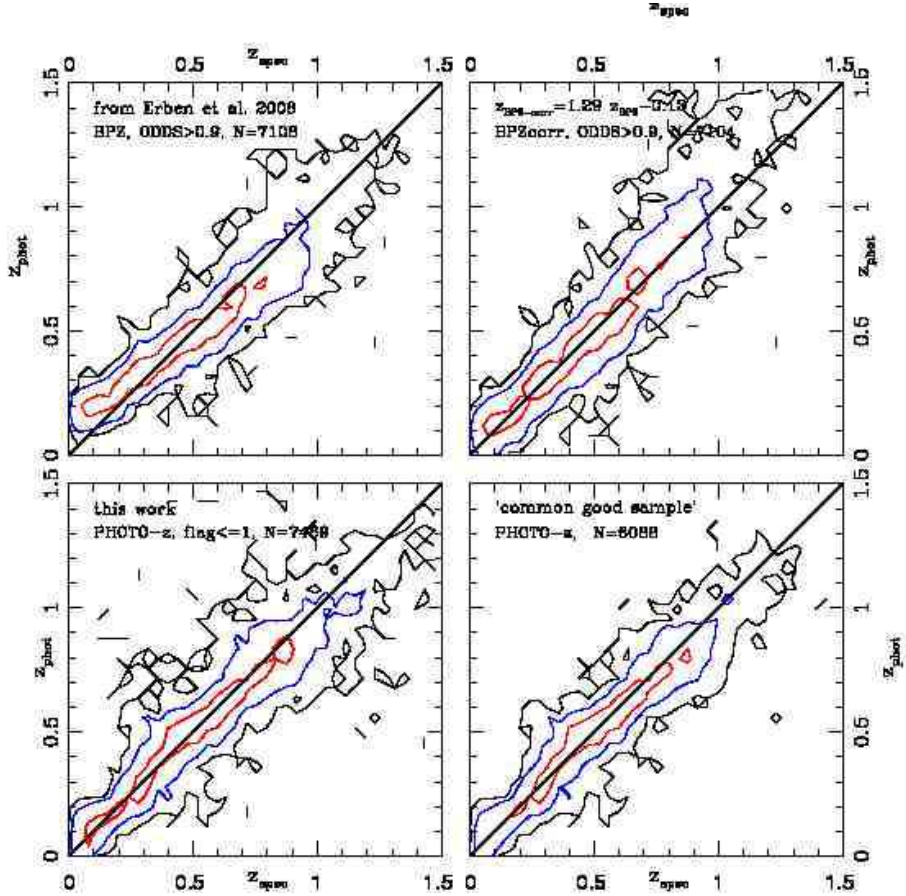} 
\caption{
Whereas the previous figure is intended to mostly show the location of
outliers, one can see the density of objects in the
spectroscopic-photometric redshift plane in this figure. The contour
levels are choosen such that they contain 99 (black), 90 (blue) and 50
(red) percent of all galaxies with spectroscopic redshifts between
zero and 1.5.  As before the upper left, upper right and lower left
panels are for the BPZ-method, the bias corrected BPZ-redshifts and
the PHOTO-z redshifts of the Bender et al. code. The density
distributions are obtained for the `good' objects for each methods
only. The exact definition can be read in Table~\ref{allz_table}., the
most relevant selection criterium can seen in the figure itself.  One
can see again the bias in the BPZ photometric redshifts, the remaining
bias at $z>0.9-1$ for the BZP-corrected case, and that PHOTO-z method
provides nearly bias free redshifts.  If one uses only objects for
which `good' photozs can be obtained with all 3 methods and shows the
result for the PHOTO-z values, one ends up with a fairly well behaved
distribution, with small dispersion and outlier rate (see details in
Table~\ref{allz_table}).
}
\label{specz_photoz_all_objects_2}
\end{figure*}
In Erben et al. \cite{erben08} we had derived photometric redshifts
with the BPZ algorithm.  In this method the `ODDS' output parameter of
the BPZ method provides a very efficient way to dissentangle likely
accurate from likely inaccurate photometric redshift. We used an odds
parameter of ODDS=$0.9$ to select reliable redshifts and displayed the
results for the reliable ones in Erben et al. \cite{erben08}.  The photo-$z$'s were
systematically overestimated at low spectroscopic redshifts and
underestimated at higher $z$. We have discussed potential origins of this 
bias already in Erben et al. \cite{erben08}.

\begin{figure*}
\centering
\includegraphics[width=18.cm]{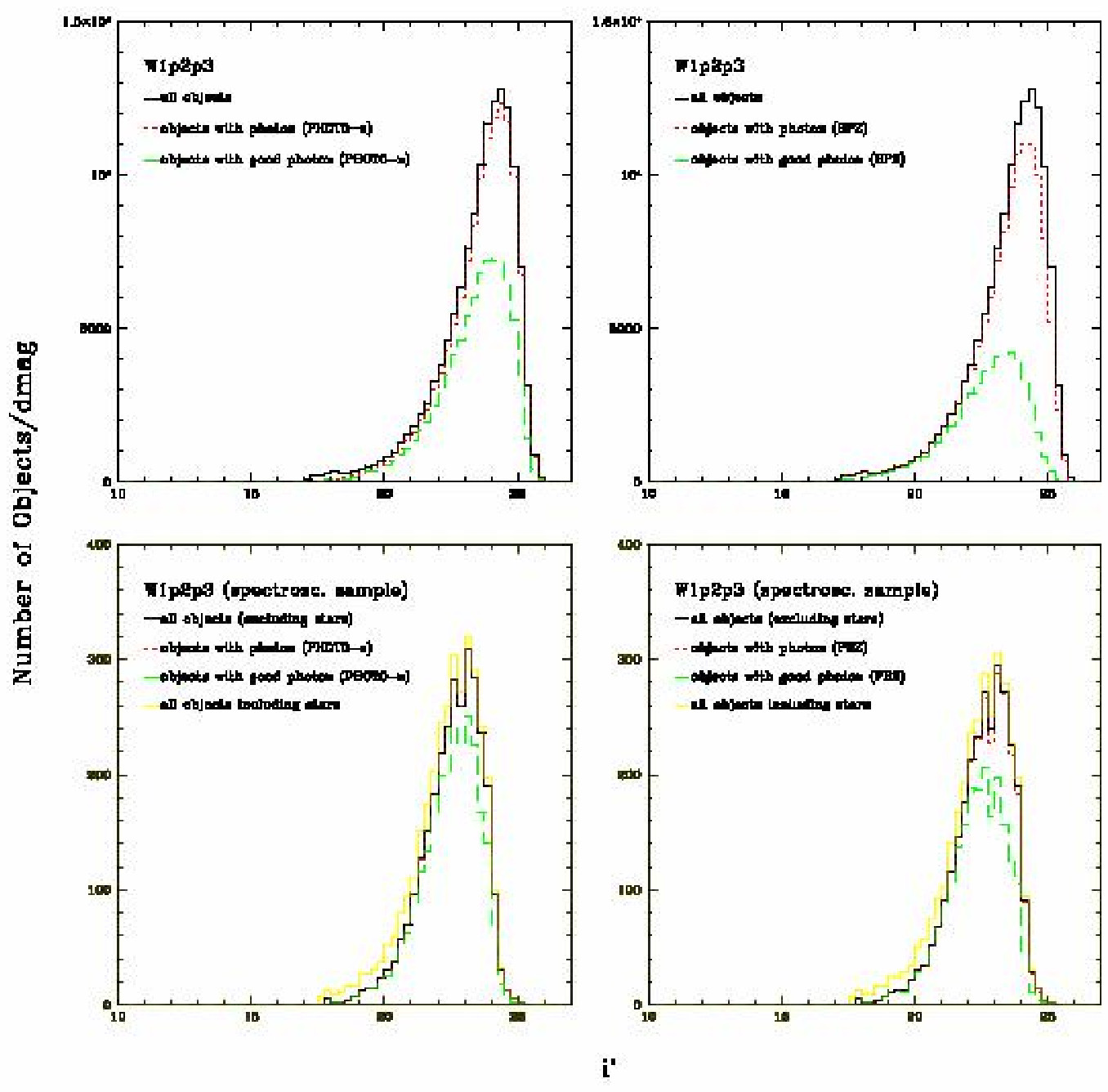} 
\caption{This figure shows the number of objects in the  \texttt{W1p2p3} field
as a function of magnitude. The upper two panels are for the
photometric catalog, the lower two panels are for the spectroscopic
catalog (from the VVDS data). Black color is used for the full
photometric and spectroscopic galaxy catalog.  (The yellow
histograms for the spectroscopic sample in addition also contain
stars).  The red histograms are for those objects, for which
photometric redshifts could be obtained, and which are not
characterized as stars by their photometry or morphology. One sees,
that the stars are properly selected out from the spectroscopic data
set, by their photometric and morphological paraemters already -- even
if the spectra were not there.  The green histograms are for those
objects which are flagged as `good' objects, i.e. which are expected
to have secure photometric redshifts. One can see, that the PHOTO-z
method (two left panels)
 can provide a larger fraction of galaxies with reliable
redshifts than the BPZ-method (two right panels).
If one considers objects to a limiting magnitude of $i'$=24.5
(which are at the same time outside any eg. bright star mask) one obtains
for 70 percent of all objects 
a `good' photometric redshift with PHOTO-z.
For, the BPZ-redshifts this is the case for 
only 54 percent of all objects (for $i'<24.5$).
}
\label{good_objects_vs_mag}
\end{figure*}

This trend has turned out to be very
stable over all investigated fields with VVDS overlap (13 fields in
total) and can be described as
\begin{equation} z_{\rm BPZ-corr} \approx 1.29 ~z_{\rm BPZ} -0.13 \quad
.
\end{equation}
\label{correct-z}
Using this empirical relation (obtained from 13 fields) between the
true redshift and the BPZ-photometric redshifts we can provide an
alternative redshift estimate, which we call $z_{\rm BPZ corr}$.
\\
We finally take all 'reliable' spectroscopic redshifts (VVDS, Deep and
SDSS) and merge them with the photometric catalogs. Requiring a
match in position within $0.56$ arcseconds distances yields a combined
catalog with photometric and spectroscopic redshifts for 9079
objects, 9130 objects and 9118 objects for the PHOTO-z, the BPZ- (from
Erben et al. 2008) and the BPZ-corrected catalogs.
Note, that the number of objects is not the same, because eg. PHOTO-z
identifies likely stars (which are then removed from the sample) and
some objects wich have small redshifts in BPZ-catalog can obtain
negative redshifts after applying the correction estimate from
equation ~\ref{correct-z}.  We show the comparison of the
spectroscopic and photometric redshifts in the \texttt{W1}, \texttt{W3} and \texttt{W4} fields in
Fig.~\ref{specz_photoz_all_objects_1}.  Black points denote all
objects in the sample, and red symbols are used, for objects, which a
method flags as having `good' or `reliable' photometric redshifts. In
the overall sample, one sees a lot of systematics, most severly for
the BPZ method (which is in the upper left panel). The relation for
bias corrected BPZ redshifts and PHOTO-z redshifts are shown in the
upper right and and lower left panels. The PHOTO-z method has fewer
systematics in the total galaxy sample shown (black points) than the
BPZ method. Fig.~\ref{specz_photoz_all_objects_1} demonstrates that
the {\rm ODDS} parameter is very helpful in sorting out outliers (see red
points in the upper panels of the same
Figure). Fig.~\ref{specz_photoz_all_objects_2} shows the density
distributions of points in the $z_{\rm true}-z_{\rm
photoz}$-space. The contours are isodensity contours, and their levels
are choosen such that they contain 99 (black), 90 (blue) and 50 (red)
percent of all objects which have true redshifts between zero and
1.5. Compared to the BPZ and BPZ-corrected redshifts, the PHOTO-z
method shows hardly any bias. We also build that sample of objects
which has `good' redshifts with both the BPZ and PHOTO-z method, and
show the comparison of true and PHOTO-z redhifts in the upper right
subpanel of Fig.~\ref{specz_photoz_all_objects_2}. These redshifts are
bias free, have an outlier rate of only $1\%$, and in quality exceed
the BPZ and BPZ-corrected redshifts according to
Table~\ref{allz_table}.
\\
This shows, that one can construct a subsample of objects with very robust 
photometric redshifts, which is useful for weak lensing studies. 
One should however try to keep
the `good'  subsample as large as possible
in order to have enough galaxies to measure the shapes.
We therefore now compare the yield of `good' objects with the PHOTO-z and 
BPZ-methods as a function of object magnitude for the photometric and 
spectroscopic sample. We use only objects from the \texttt{W1p2p3}-field (which are
located in an area not flagged having potentially unreliable photometry, i.e.
objects having a flag of zero in the photometric catalog) and show 
results in Fig.~\ref{good_objects_vs_mag}.
The histogram for the number
of objects with a given magnitude in the \texttt{W1}-field is shown in black in the 
upper two panels of this figure. The same histogram for objects which have
spectroscopic data (and are not stars) in the  \texttt{W1}-field is shown in black in the lower two panels. All objects (including stars) are shown as yellow histogram.
One can see, that there are fairly many stars at the bright end.
The histograms for those objects for which photometric redshifts 
could be derived and which are not classified as stars in terms of 
morphology or SED  
are shown in red, and those which have `good' photometric redshifts are 
shown in green. The left panels are for PHOTO-z redshifts, the right panels
are for BPZ-redshifts. One can see, that the yield of `good' objects is quite 
complete for the spectroscopic and photometric sample for both the PHOTO-z and 
BPZ-method up to $i'=22$. 
For fainter magnitudes the BPZ-method is less complete: 
one obtains good redshifts for objects brighter than 
$i'$=24.5 only for 54 percent of all objects using BPZ.
For the PHOTO-z method, this ratio equals 70 percent.

We are aware that the fraction of galaxies which have `good' photometric 
redshifts (low clipped dispersion and low outlier rate) should be increased, 
or become `identical' to the original sample.  This can be achieved by 
adding NIR data, by improving our SED-templates, and probably more important
by making our photometry  (including the convolution to the same
PSF)  more accurate. This will be subject of a further study.
\begin{table*}
\caption{Photometric redshifts of PHOTO-z, BPZ and BPZ-corr in the  \texttt{W1},  \texttt{W3} and  \texttt{W4} fields 
are  compared to the spectroscopic sample consisting of VVDS, DEEP2, SDSS DR6 spectroscopic data. In this table, 
$\sigma$ denotes the `true' dispersion, without clipping outliers, and $\sigma_{clip}$  denotes the width of 
the distributions after clipping $\Delta z_{\rm photoz} >0.25* (1+z_{\rm photoz})$ outliers 
(as defined in equation~\ref{sigclip} and introduced by Ilbert et al. 2006).
}
\label{tab:compcheck}
\centering
\renewcommand{\footnoterule}{} 
\begin{tabular}{llllllll}
\hline 
Code & Sample: CFHTLS-W1, & $N_{zspec}$
                                                                                        & Median-error & Mean-error &  $\sigma$ &  $\sigma_{clip}$   & $\eta$ [\%] \\
\hline
PHOTO-z	     &	all objects         
                                                                          &	9079	&	-0.0036	 &	0.0065	  &	0.158	&      	0.040 &	 5.7$\%$   \\	
PHOTO-z	     &	good PHOTO-z objects
                                                                          &     7469	&	-0.0025	 &      0.0019	  &     0.071	&      	0.038 &  2.8$\%$  \\		
PHOTO-z	     &	common good objects 
                                                                          &	6088	&	-0.0025	 &      0.0001  &     0.053	&      	0.037 &  1.0$\%$\\	
\hline                                                                                                                                                 
BPZ	     &	all objects      
                                                                 	  &     9130	&	-0.0005 &      0.0257    &      0.194	&      	0.056 &  10.5$\%$	\\	
BPZ	     &	good BPZ objects 
                                                  			  &     7108	&	-0.0017	 &      0.0013   &      0.066	&      	0.054 &   2.7$\%$	\\	
BPZ	     &	common good objects 
                                                                          &	6088	&	-0.0050	 &      0.0000 &      0.062 &      	0.056 &   1.5$\%$\\	
\hline                                                                                                                                                 
${\rm BPZ_{\rm corr}}$  &	all objects 
                                                                          &     9118	&	+0.0076	 &       0.0414   &      0.241	&      	0.051 &  11.0$\%$	\\	
${\rm BPZ_{\rm corr}}$  &	good BPZ objects 
                                                 			  &     7104	&	+0.0024	 &       0.0055  &      0.065	&      	0.047 &   2.7$\%$	\\	
${\rm BPZ_{\rm corr}}$  &	common good objects 
                                                                          &	6088	&	+0.0063	 &       0.0098  & 	 0.060 &      	0.048 &   1.8$\%$\\	
\hline
\end{tabular}
\begin{flushleft}
The samples are defined as follows: \\
all PHOTO-z objects:   {$ z_{\rm spec}>0,z_{\rm PHOTO-z}>0 $} \\
good PHOTO-z objects:  {$z_{\rm spec}>0,z_{\rm PHOTO-z}>0,flag_{\rm PHOTO-z}<=1$} \\
common good objects:   {$z_{\rm spec}>0,z_{\rm PHOTO-z}>0,z_{\rm BPZ}>0,z_{\rm BPZ-corr}>0,flag_{\rm PHOTO-z}<=1, ODDS_{\rm BPZ}>0.9$} \\
	all BPZ- objects:       { $z_{\rm spec}>0,z_{\rm BPZ}>0$} \\
	good BPZ objects:   {$z_{\rm spec}>0,z_{\rm BPZ}>0,ODDS_{\rm BPZ}>0.9$} \\
	all BPZ-corr objects:   { $z_{\rm spec}>0,z_{\rm BPZ-corr}>0$} \\
	good BPZ-corr objects:   {$z_{\rm spec}>0,z_{\rm BPZ-corr}>0,ODDS_{\rm BPZ}>0.9$} \\
\end{flushleft}
\label{allz_table}
\end{table*}
\begin{figure}
\centering
\includegraphics[width=8.5cm]{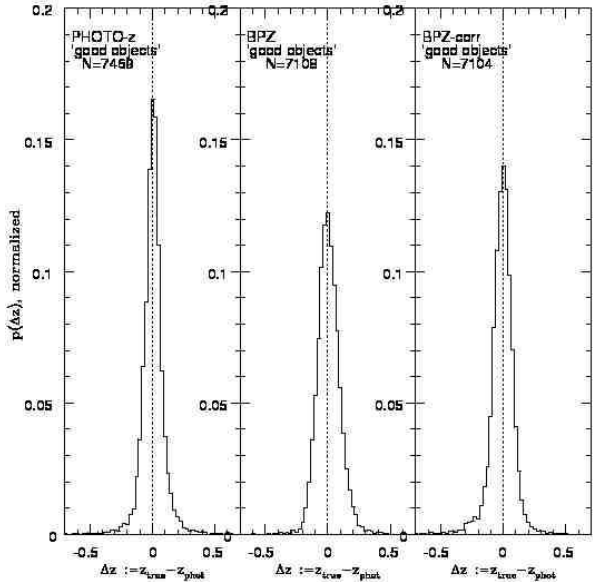} 
\caption{This figure shows normalized histograms for the photometric
redshift error, $\Delta z = z_{\rm true}-z_{\rm photoz}$ for the
PHOTO-z method (left panel), for the BPZ-method (middle panel) and for
the BPZ-(bias) corrected catalogs (right panel), using the  \texttt{W1},  \texttt{W2}, and 
 \texttt{W3} field and all reliable spectroscopic VVDS, DEEP2 and SDSS redshift data. 
Redshift biases as
seen in the BPZ catalog don't show up in this histogram (which
combines all spectroscopic redshifts up to $z=4.5$) since only the
histograms within small redshift slices are shifted relative to
$\Delta z=0$. One can see, however, that the PHOTO-z redshifts have
more objects with redshifts very close to the true redshift relatively
to the other two methods (note, that these histograms are
normalized). In addition to that, the PHOTO-z method provides the
largest amount of galaxies in its `good' sample.
}
\label{specz_photoz_all_objects_3}
\end{figure}

\section{Summary}
\label{sec:sum}
We tested the performance of the photometric redshift code of Bender
et al. (2001) with the CFHT-MegaCam filter system using the CFHTLS \texttt{D1}
data. The comparison of our photometric redshift results with
spectroscopic data and photometric redshift results from Ilbert et
al. \cite{ilbert06} shows: our performance is very close to that of
Ilbert et al. \cite{ilbert06}, although we use only about half of the optical bands
(and no NIR data).
This makes us confident 
that results we then derived for the CFHTLS-Wide fields are reliable.  

We analyzed the CFHTLS-Wide data and showed that the colors of stars
can be measured accurately enough and that the throughput of the
system is known well enough to allow relative zero point calibration
for the $g'r'i'z'$ bands using the colors predicted from Pickles
libary stars.  We could not match the color-color diagrams on 
the Pickles-libary-star colors when the $u^*$-band is involved. We
nevertheless found out, that calibrating zeropoints such that the
$u^*-g'$ colors approximately match at the red end, gives results which 
describe the galaxy colors correctly.  In this case the
likely reason for the mismatch would be that Pickles stars are metal
enriched and do not show the UV excess of metal poor halo stars.

After the improved relative zeropoint calibration we derived
photometric redshifts for 2.5 million of galaxies. We identified
galaxies with likely inaccurate redshifts as those which have formally
a large photometric redshift error (as provided by the code) and which
have SEDs which are likely to be mismatched with another SED at 
different redshifts. By flagging those objects we end up with a sample
of galaxies with fairly precise photometric redshifts. We investigated
the redshift accuracy as a function of brightness and SED-type, and
find similar results (in numbers and in trend) as Ilbert et al. \cite{ilbert06} for
CFHTLS-Deep data set.

The overall photometric redshift precision was quantified by comparing
all  \texttt{W1},  \texttt{W3} and  \texttt{W4} photometric redshifts to VVDS, DEEP2 and SDSS
spectra. We then also investigated the BPZ-redshifts from Erben et al. \cite{erben08}. 
These redshifts are biased, and can be corrected according to
$z_{\rm BPZ-corr}= 1.29 ~z_{\rm BPZ}-0.13$. This correction slightly
overcorrects at redshifts larger than 1.  We then analyzed all three
photometric redshifts samples (PHOTO-z, BPZ, BPZ-corr) in more detail:
Taking all objects (irrespective of photoz quality flags) the outlier-rate
varies between 10 percent (BPZ/BPZ-corr) and 6 percent (PHOTO-z). If
we select only good objects (using photometric redshift errors and SED
types for the PHOTO-z method and using the ODDS parameter for the BPZ
method) we can reduce the outlier rate to about 2.7 percent for all
three catalogs. The width of the distributions (after clipping) then
becomes 0.038, 0.054 and 0.057 for the PHOTO-z, the BPZ and the
BPZ-corr catalogs.  The width of the (unclipped) distribution is
0.071 (PHOTO-z) and 0.066 (BPZ/BPZ-corr).

Finally we consider only galaxies which are classified as `reliable'
objects in all three catalogs, and investigate the photometric
redshift quality for this `common sample'.  We indeed can reduce the
outlier rate to 1 percent (PHOTO-z) and 1.5 to 1.8 percent for the
BPZ-versions. The width of the distributions (after clipping) then
becomes 0.037, 0.056 and 0.058 for the PHOTO-z, the BPZ and the
BPZ-corr catalogs.  The width of the (unclipped) distribution is
0.05 (PHOTO-z) and 0.06 (BPZ/BPZ-corr).

We conclude that this common sample defines a high quality redshift
sample, which has (in the case of PHOTO-z) no bias and a very low
outlier rate. This sample is ideally suited for weak lensing analysis
like growth of cosmic shear and in particular the shear ratio test
behind clusters of galaxies. Since these `good' redshift samples 
include several selections steps, and are fairly incomplete at faint magnitudes
we don't recommend this sample to be taken for galaxy evolution studies in 
general. 

The original sample (all galaxies) can be taken for that, which however 
requires to understand the impact of outliers (eg. on derived luminosity 
functions, galaxy colors as a function of redshift and environment density);

We are currently working  on increasing the `good' sample, 
or decreasing the outlier rate of the `remaining' sample, to finally unite
them to one again. Goal is to obtain a `complete sample' with outlier rates
as low as 2 percent. 
This requires a more detailed study of photometric 
calibration, improved convolution for more precise aperture colors, improved 
SEDs, potentially varying priors, and including further colors where 
possible. 

We provide these catalogs (with future updates and extensions) on
request.
%
%
%
\begin{acknowledgements}
We are greatful to the Terapix consortium for developing and providing tools for the handling of large CCD images in general, and for the image processing and pipeline software for MegaCam, and finally for the production of photometrically and astrometically corrected images within the CFHTLS survey. This work uses images which have been compared to the CFHTL T0003 release when deriving the photometric and astrometric accuracy.

We acknowledge use of the Canadian Astronomy Data Centre, which is operated by the Dominion Astrophysical Observatory for the National Research Council of Canada's Herzberg Institute of Astrophysics. 

We thank Y. Mellier and J. Coupon for the friendly spirit in which the
Munich and Paris teams independently worked on their photometric redshifts. \\

This work was supported by the DFG Sonderforschungsbereich 375 "Astro-Teilchenphysik", the DFG priority program 1177 (Se1038), TRR33 "The Dark Universe" and the DFG `Cluster of Excellence on the Origin and Structure of the Universe''. 
\\
We all, in particular M. L., thank the European Community for the support by
the Marie Curie research training network "DUEL". 
M.L. further thanks the University of Bonn and the University of 
British Columbia for hospitality. 
\end{acknowledgements}
%
%
%

%
%
%
\begin{appendix}
\section{Details on the photometric redshift redshift catalog}
Using the notation introduced in Erben et al. \cite{erben08}, we briefly explain the most important \texttt{FITS} keys in the multi-color catalogs in the Table~\ref{tab:keys} .\\
\begin{table*}
\begin{minipage}[t]{\textwidth}
\caption{Description of the most important \texttt{FITS} keys in the \CARS multi-color catalogs. The \texttt{ASCII} catalog version contains one aperture magnitude at a diameter of $1\myarcsec 86$.}
\label{tab:keys}
\centering
\renewcommand{\footnoterule}{}  
\begin{tabular}{lll}
\hline
\hline
key name & description & measured on\\
\hline
\texttt{SeqNr} & Running object number & -\\
\texttt{ALPHA\_J2000} & Right ascension & unconvolved $i$-band image\\
\texttt{DELTA\_J2000} & Declination & unconvolved $i$-band image\\
\texttt{Xpos} & x pixel position & unconvolved $i$-band image\\
\texttt{Ypos} & y pixel position & unconvolved $i$-band image\\
\texttt{FWHM\_WORLD} & FWHM assuming a Gaussian core & unconvolved $i$-band image\\
\texttt{FLUX\_RADIUS} & half-light-radius & unconvolved $i$-band image\\
\texttt{A\_WORLD} & profile RMS along major axis & unconvolved $i$-band image\\
\texttt{B\_WORLD} & profile RMS along minor axis & unconvolved $i$-band image\\
\texttt{THETA\_WORLD} & position angle & unconvolved $i$-band image\\
\texttt{Flag} & \SExtractor extraction flags & unconvolved $i$-band image\\
\texttt{CLASS\_STAR} & star-galaxy classifier & unconvolved $i$-band image\\
\texttt{MAG\_AUTO} & total $i$-band magnitude & unconvolved $i$-band image\\
\texttt{MAGERR\_AUTO} & total $i$-band magnitude error & unconvolved $i$-band image\\
\texttt{MAG\_ISO\_x\footnote{$\mathrm{x} \in \left[ u,g,r,i,z\right]$}} & isophotal magnitude in x-band & PSF-equalised x-band image\\
\texttt{MAGERR\_ISO\_x} & isophotal magnitude error in x-band & PSF-equalised x-band image\\
\texttt{MAG\_APER\_x} & aperture magnitude vector in x-band & PSF-equalised x-band image\\
\texttt{MAGERR\_APER\_x} & aperture magnitude error vector in x-band & PSF-equalised x-band image\\
\texttt{MAG\_LIM\_x} & limiting magnitude in x-band & unconvolved x-band image\\
\texttt{FLUX\_ISO\_x} & isophotal flux in x-band & PSF-equalised x-band image\\
\texttt{FLUXERR\_ISO\_x} & isophotal flux error in x-band & PSF-equalised x-band image\\
\texttt{FLUX\_APER\_x} & aperture flux vector in x-band & PSF-equalised x-band image\\
\texttt{FLUXERR\_APER\_x} & aperture flux error vector in x-band & PSF-equalised x-band image\\
\hline
\texttt{Z1\_PHOT} & photometric redshift best-fit SED & -\\
\texttt{ERR\_Z1\_PHOT} & error of photometric redshift best-fit SED & -\\
\texttt{SED\_TYPE }& SED type & -\\
\texttt{Flag\_PHOT} & global photometric redshift flag key \footnote{$1=$~object FHWM$<$PSF, $2=\triangle z > 0.25 * (1+z)$, $4=$~ext. flag~(\SExtractor,
absolute photometry) for object, $8=$~SED rejected, \\$16=$~star SED, $32=$~no photometric redshift, and combinations.} & -\\
\hline
\texttt{MASK} & global mask key\footnote{0 for objects inside masks and 1 otherwise} & -\\
\end{tabular}
\end{minipage}
\end{table*}
\end{appendix} 
\end{document}